\documentclass[conference,letterpaper]{IEEEtran}
\IEEEoverridecommandlockouts
\usepackage[utf8]{inputenc} 
\usepackage[T1]{fontenc}
\usepackage{url}
\usepackage{ifthen}
\usepackage{cite}
\usepackage[cmex10]{amsmath} 

\usepackage{stmaryrd}
\usepackage[cmex10]{mathtools}
\usepackage{amssymb}
\usepackage{tikz}
\usepackage{subfig}
\usepackage{bm}
\usepackage{mathrsfs}

\newtheorem{definition}{Definition}
\newtheorem{example}{Example}
\newtheorem{proposition}{Proposition}
\newtheorem{theorem}{Theorem}
\newtheorem{lemma}{Lemma}
\newtheorem{remark}{Remark}

\newcommand*{\boldone}{\text{\usefont{U}{bbold}{m}{n}1}}

\def\DrawTikz{}

\begin{document}
\title{Generalized Orthogonal de Bruijn and Kautz Sequences} 

\author{%
 \IEEEauthorblockN{Yuan-Pon Chen, Jin Sima and Olgica Milenkovic}
 \IEEEauthorblockA{Department of Electrical an Computer Engineering\\
                   University of Illinois Urbana-Champaign\\
                   Champaign, IL, USA\\
                   Email: \{yuanpon2, jsima, milenkov\}@illinois.edu}
\thanks{
Parts of the work have been submitted to the IEEE Symposium on Information Theory, Ann Arbor, 2025~\cite{chen2025generalized}. This extension contains added proofs for results on orthogonal de Bruijn sequences and a completely new section on orthogonal Kautz sequences.
}
}

\maketitle

\begin{abstract}
A de Bruijn sequence of order $k$ over a finite alphabet is a cyclic sequence with the property that it contains every possible $k$-sequence as a substring exactly once. Orthogonal de Bruijn sequences are collections of de Bruijn sequences of the same order, $k$, satisfying the joint constraint that every $(k+1)$-sequence appears as a substring in at most one of the sequences in the collection. Both de Bruijn and orthogonal de Bruijn sequences have found numerous applications in synthetic biology, although the latter remain largely unexplored in the coding theory literature. Here we study three relevant practical generalizations of orthogonal de Bruijn sequences where we relax either the constraint that every $(k+1)$-sequence appears exactly once, or that the sequences themselves are de Bruijn rather than balanced de Bruijn sequences. We also provide lower and upper bounds on the number of fixed-weight orthogonal de Bruijn sequences. The paper concludes with parallel results for orthogonal nonbinary Kautz sequences, which satisfy similar constraints as de Bruijn sequences except for only being required to cover all subsequences of length $k$ whose maximum runlength equals to one.
\end{abstract}

\IEEEdisplaynontitleabstractindextext

\section{Introduction}

De Bruijn sequences~\cite{de1975acknowledgement,van1987circuits} are combinatorial objects that have found many practical applications, ranging from pseudorrandomess generation,  hashing and lookup table design to DNA assembly and molecular data storage~\cite{DNA_profiles_Kiah}. The utility of de Bruijn sequences of order $k$ stems from the fact that they have the property that they cover all $k$-sequences over a finite alphabet as substrings exactly once. De Bruijn sequences have  been further generalized to include \emph{balancing constraints}~\cite{Balanced_dB_Etzion}, in which case every $k$-sequence is allowed to appear $\ell \geq 1$ or at most $\ell$ times, or general substring constraints as described in~\cite{dB_Fix_W_Ruskey}. In addition, they have been modified to accommodate other practical constraints, such as runlength bounds, in which case the sequences are known as \emph{Kautz sequences}~\cite{rosenfeld2002enumerating}.  

Another interesting extension of the concept of de Bruijn sequences is \emph{orthogonal de Bruijn sequences}, introduced in~\cite{Ortho_dB_Skiena} for the purpose of optimizing experimental designs in synthetic biology\footnote{Orthogonal de Bruijn sequences were independently studied in the mathematics literature~\cite{Compatible_Euler_Fleischner,Arc_Disjoint_Rowley_Bose} under the name of arc-disjoint de Bruijn cycles.}. In a nutshell, orthogonal de Bruijn sequences are de Bruijn sequences of order $k$ that satisfy the joint (cross) property that every $(k+1)$-sequence appears in at most one of the sequences in the collection. The de Bruijn property of the sequences is used to ensure both the \emph{diversity of DNA sequence binding probes} of length $k$ and the shortest sequence length property under the diversity constraint (since DNA strings have to be synthesized for testing and since the cost of synthesis prohibits the use of long strings). Interestingly, the orthogonality constraint aims to reduce undesired cross-hybridization of longer probes designed to target only one of the sequences; although in the definition the constraint is imposed on length-$(k+1)$ substrings, other constrained substring lengths (such as $2k$) are equally relevant. From the perspective of DNA-based data storage, orthogonal de Bruijn sequences are relevant for multistage primer-based random access~\cite{tabatabaei2015rewritable}. There, groups of strings sharing a common $k$-substring can be accessed together, and then further partitioned into subgroups as needed using more specialized primers that are not shared by the whole group (say, using primers of length $k+s$, $s>0$).     
A drawback of orthogonal de Bruijn sequences is that their number is strongly restricted by the alphabet size. One can increase the orthogonal collection by relaxing the notion of orthogonality, as described in Section 3 (following the preliminaries of Section 2). There we study orthogonal de Bruijn sequences in which $(k+1)$-sequences are allowed to appear at most $\ell$ times, with $\ell\geq 1$. The main result is that the number of generalized orthogonal de Bruijn sequences scales with $\ell$. 

Another practical issue with orthogonal de Bruijn sequences is that each $k$-sequence has only one context in which it appears in each of the sequences. To increase the number of different contexts, one can examine $b$-balanced de Bruijn sequences, in which each $k$-sequence is allowed to appear
exactly
$b$ times~\cite{Balanced_dB_Etzion}. In this case, one can investigate the (new) notion of orthogonality of balanced de Bruijn sequences, as outlined in Section 4. For ease of synthesis, it is desirable to maintain composition/weight constraints for the DNA sequences, resulting in counting and construction problems pertaining to fixed-weight (fixed composition) de Bruijn sequences, introduced and studied in Section 5.

We conclude our exposition with a review of Kautz and orthogonal Kautz sequences and the introduction of the concepts of balanced and fixed-weight Kautz and orthogonal Kautz sequences, as described in Section 6. The relevance of the added runlength constraint is that the sequences avoid what is known as \emph{homopolymers} of DNA symbols, which are known to cause both DNA synthesis and sequencing errors. This is particularly the case for nanopore sequencers, as first described and experimentally evaluated in~\cite{yazdi2017portable}.

\section{Preliminaries}
We start by reviewing relevant concepts and definitions pertaining to (orthogonal) de Bruijn sequences and graphs.
\begin{definition}\label{def:deB_seq}
    Let $k\geq 1$ and let $\mathcal{A}$ be an alphabet of size $\sigma\geq 2$.
    A circular sequence $\mathbf{s}=(s_0,s_1,\ldots,s_{\sigma^k-1})$ is called
        a \emph{$(\sigma,k)$-de Bruijn sequence}
        if each length-$k$ sequence in
        $\mathcal{A}^k$
        appears as a
        circular
        substring of $\mathbf{s}$ exactly once.
    More specifically,
        for each length-$k$ sequence $\mathbf{t}=(t_0,\ldots,t_{k-1})\in
        \mathcal{A}^k
        $,
        there is a unique index $i\in [0, \sigma^k-1]$ s.t. 
        $(s_i,s_{i+1\mod \sigma^k},\ldots,s_{i+k-1 \mod \sigma^k})=\mathbf{t}$. We do not distinguish sequences that are circular shifts of each other.
\end{definition}

    Fundamental for the study of de Bruijn sequences is the notion of a \emph{de Bruijn graph} of order $k$ and alphabet size $\sigma$,
        denoted by $G_{\sigma,k}$.
        A de Bruijn graph is a directed graph $(V,A)$ with vertex set $V=\mathcal{A}^{k-1}$
        and arc set
        \begin{align}
            A =
            \{
            ((s_0,s_1,\ldots,s_{k-2}),(s_1,\ldots,s_{k-2},s_{k-1}))
            \nonumber\\
            ~|~ s_0,\ldots,s_{k-1}\in\mathcal{A}
            \}.
        \end{align}
    In words,
        there exists an arc from $v_1$ to $v_2$ iff 
        the length-$(k-2)$ suffix of $v_1$
        is the same as
        the length-$(k-2)$ prefix of $v_2$.

\begin{definition}\label{def:ortho}
    A collection of $(\sigma,k)$-de Bruijn sequences $\mathcal{C}=\{\mathbf{s}_1,\ldots,\mathbf{s}_M\}$
    is called
    \emph{orthogonal}
    if each length-$(k+1)$ sequence in $\mathcal{A}^{k+1}$
    appears at most once in   $\mathcal{C}$.
\end{definition}

 It is clear that there exists a 1-1 correspondence between
        $(\sigma,k)$-de Bruijn sequences
        and
        Eulerian circuits (i.e., closed walks that traverse each edge exactly once) in $G_{\sigma,k}$.
    Furthermore, a length-$(k+1)$ string $(s_0,\ldots,s_k)$
        appears in a $(\sigma,k)$-de Bruijn sequence iff 
        the corresponding Eulerian circuit
        traverses
        from $\mathbf{v}_1$ to $\mathbf{v}_2$ to $\mathbf{v}_3$,
        where
        $\mathbf{v}_1 = (s_0,\ldots,s_{k-2})$,
        $\mathbf{v}_2 = (s_1,\ldots,s_{k-1})$, and
        $\mathbf{v}_3 = (s_2,\ldots,s_{k})$.

    Similarly, there exists a 1-1 correspondence between
        $(\sigma,k)$-de Bruijn sequences
        and
        Hamiltonian cycles (i.e., closed walks that traverse every vertex exactly once) in  $G_{\sigma,k+1}$.
    The length-$(k+1)$ string $(s_0,\ldots,s_k)$
        appears in a $(\sigma,k)$-de Bruijn sequence
        iff
        the corresponding Hamiltonian cycle
        traverses from $(s_0,\ldots,s_{k-1})$
        to
        $(s_1,\ldots,s_{k})$. 

The relevance of de Bruijn graphs $G_{\sigma,k+1}$ for the analysis of orthogonal de Brujin sequences comes from the fact that certain arc-constrained Hamiltonian cycles correspond to orthogonal sequences. In that context, it was shown in~\cite{Ortho_dB_Skiena} that for $\sigma \geq 3$, the number of orthogonal $(\sigma,k)$-de Bruijn sequences is bounded between $\lfloor \sigma /2 \rfloor$ and  $\sigma-1$.       

\begin{example}
Let $k=2$ and $\mathcal{A}=\{0,1,2\}$.
The circular sequence
    $\mathbf{s}=012002211$
    is a $(3,2)$-de Bruijn sequence
    since 
    its length-$2$ substrings are
    $01,12,20,00,02,22,21,11,10$ -- all the length-$2$ sequence over $\mathcal{A}$, 
    without repetition.
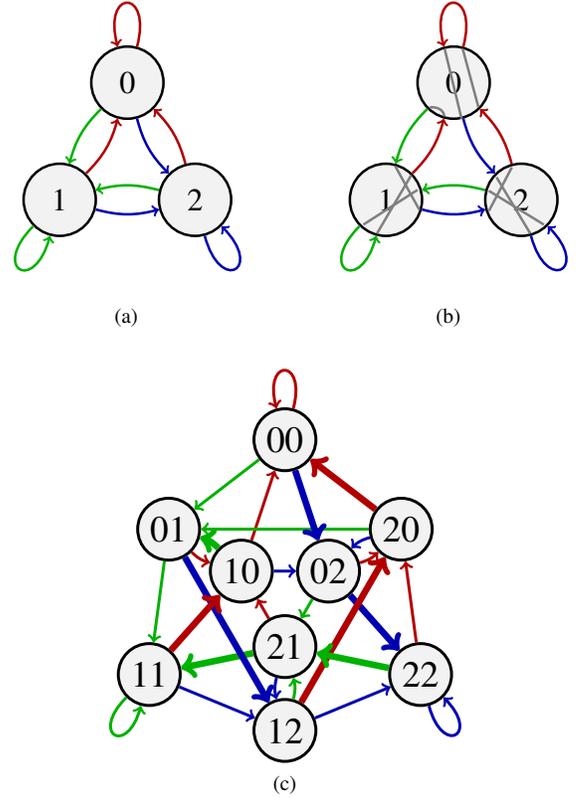
\begin{figure}
    \centering
    \subfloat[][]{
        \scalebox{0.6}{
            \ifdefined\DrawTikz
                \newcommand{\LD}{\langle}
\newcommand{\RD}{\rangle}
\begin{tikzpicture}[scale=1]
\node[draw, circle, color=black, fill=black!5, ultra thick,minimum size=1.6cm] (0) at (0,1.5*1.732) {\huge 0};
\node[draw, circle, color=black, fill=black!5, ultra thick,minimum size=1.6cm] (1) at (-1.5,0) {\huge 1};
\node[draw, circle, color=black, fill=black!5, ultra thick,minimum size=1.6cm] (2) at (1.5,0) {\huge 2};

\draw[->, ultra thick, color = green!70!black] (0) to [out=225,in=75] (1);
\draw[->, ultra thick, color = blue!70!black] (1) to [out=-15,in=195] (2);
\draw[->, ultra thick, color = red!70!black] (2) to [out=105,in=-45] (0);

\draw[->, ultra thick, color = blue!70!black] (0) to [out=-75,in=135] (2);
\draw[->, ultra thick, color = green!70!black] (2) to [out=165,in=15] (1);
\draw[->, ultra thick, color = red!70!black] (1) to [out=45,in=255] (0);

\draw[->, ultra thick, color = red!70!black] (0) to [out=75,in=105,loop] (0);
\draw[->, ultra thick, color = green!70!black] (1) to [out=225,in=255,loop] (1);
\draw[->, ultra thick, color = blue!70!black] (2) to [out=285,in=315,loop] (2);





\end{tikzpicture}%
            \fi
        }
        \label{subfig:dB_3_2_original}
    }
    \subfloat[][]{
        \scalebox{0.6}{
            \ifdefined\DrawTikz
                \newcommand{\LD}{\langle}
\newcommand{\RD}{\rangle}
\begin{tikzpicture}[scale=1]
\node[draw, circle, color=black, fill=black!5, ultra thick,minimum size=1.6cm] (0) at (0,1.5*1.732) {\huge 0};
\node[draw, circle, color=black, fill=black!5, ultra thick,minimum size=1.6cm] (1) at (-1.5,0) {\huge 1};
\node[draw, circle, color=black, fill=black!5, ultra thick,minimum size=1.6cm] (2) at (1.5,0) {\huge 2};

\draw[->, ultra thick, color = green!70!black] (0) to [out=225,in=75] (1);
\draw[->, ultra thick, color = blue!70!black] (1) to [out=-15,in=195] (2);
\draw[->, ultra thick, color = red!70!black] (2) to [out=105,in=-45] (0);

\draw[->, ultra thick, color = blue!70!black] (0) to [out=-75,in=135] (2);
\draw[->, ultra thick, color = green!70!black] (2) to [out=165,in=15] (1);
\draw[->, ultra thick, color = red!70!black] (1) to [out=45,in=255] (0);

\draw[->, ultra thick, color = red!70!black] (0) to [out=75,in=105,loop] (0);
\draw[->, ultra thick, color = green!70!black] (1) to [out=225,in=255,loop] (1);
\draw[->, ultra thick, color = blue!70!black] (2) to [out=285,in=315,loop] (2);







\draw[gray, ultra thick] (-1.29,0.74) -- (-0.76,-0.22);
\draw[gray, ultra thick] (0.78,-0.25) -- (1.29,0.72);
\draw[gray, ultra thick] (0.55,2.05) -- (0.2,3.35);
\draw[gray, ultra thick] (-0.2,3.35) -- (0.175,1.8);
\draw[gray, ultra thick] (0.95,0.5) -- (1.7,-0.75);
\draw[gray, ultra thick] (2,-0.55) -- (0.78,0.2);
\draw[gray, ultra thick] (-0.8,0.22) -- (-2,-0.55);
\draw[gray, ultra thick] (-1.7,-0.75) -- (-0.9375,0.55);
\draw[gray, ultra thick] (-0.21,1.85) to[out=90,in=7.5] (-0.525,2.05);

\end{tikzpicture}%
            \fi
        }
        \label{subfig:dB_3_2_012002211}
    }\\
    \subfloat[][]{
        \scalebox{0.7}{
            \ifdefined\DrawTikz
                \begin{tikzpicture}[scale=0.85]
\node[draw, circle, color=black, fill=black!5, ultra thick,minimum size=1cm] (00) at (90:3.5cm) {\huge 00};
\node[draw, circle, color=black, fill=black!5, ultra thick,minimum size=1cm] (11) at (210:3.5cm) {\huge 11};
\node[draw, circle, color=black, fill=black!5, ultra thick,minimum size=1cm] (22) at (330:3.5cm) {\huge 22};
\node[draw, circle, color=black, fill=black!5, ultra thick,minimum size=1cm] (01) at (150:3cm) {\huge 01};
\node[draw, circle, color=black, fill=black!5, ultra thick,minimum size=1cm] (12) at (270:3cm) {\huge 12};
\node[draw, circle, color=black, fill=black!5, ultra thick,minimum size=1cm] (20) at (30:3cm) {\huge 20};
\node[draw, circle, color=black, fill=black!5, ultra thick,minimum size=1cm] (10) at (150:1.125cm) {\huge 10};
\node[draw, circle, color=black, fill=black!5, ultra thick,minimum size=1cm] (21) at (270:1.125cm) {\huge 21};
\node[draw, circle, color=black, fill=black!5, ultra thick,minimum size=1cm] (02) at (30:1.125cm) {\huge 02};


\draw[->, ultra thick, color = red!70!black] (00) to [out=75, in=105, loop] (00);
\draw[->, ultra thick, color = red!70!black] (10) -- (00);
\draw[->, line width = 0.125cm, color = red!70!black] (20) -- (00);

\draw[->, ultra thick, color = green!70!black] (01) -- (11);
\draw[->, ultra thick, color = green!70!black] (11) to[out=225, in=255, loop] (11);
\draw[->, line width = 0.125cm, color = green!70!black] (21) -- (11);

\draw[->, line width = 0.125cm, color = blue!70!black] (02) -- (22);
\draw[->, ultra thick, color = blue!70!black] (12) -- (22);
\draw[->, ultra thick, color = blue!70!black] (22) to[out=285, in=315, loop] (22);

\draw[->, ultra thick, color = green!70!black] (00) -- (01);
\draw[->, line width = 0.125cm, color = green!70!black] (10) to[out=135,in=-15] (01);
\draw[->, ultra thick, color = green!70!black] (20) -- (01);

\draw[->, line width = 0.125cm, color = blue!70!black] (01) -- (12);
\draw[->, ultra thick, color = blue!70!black] (11) -- (12);
\draw[->, ultra thick, color = blue!70!black] (21) to[out=255,in=105]  (12);

\draw[->, ultra thick, color = red!70!black] (02) to[out=15,in=225] (20);
\draw[->, line width = 0.125cm, color = red!70!black] (12) -- (20);
\draw[->, ultra thick, color = red!70!black] (22) -- (20);

\draw[->, line width = 0.125cm, color = blue!70!black] (00) -- (02);
\draw[->, ultra thick, color = blue!70!black] (10) -- (02);
\draw[->, ultra thick, color = blue!70!black] (20) to[out=195,in=45] (02);

\draw[->, ultra thick, color = green!70!black] (02) -- (21);
\draw[->, ultra thick, color = green!70!black] (12) to[out=75,in=285] (21);
\draw[->, line width = 0.125cm, color = green!70!black] (22) -- (21);

\draw[->, ultra thick, color = red!70!black] (01) to[out=-45,in=165] (10);
\draw[->, line width = 0.125cm, color = red!70!black] (11) -- (10);
\draw[->, ultra thick, color = red!70!black] (21) -- (10);





\end{tikzpicture}%
            \fi
        }
        \label{subfig:dB_3_3_012002211}
    }
    \caption{
        \protect\subref{subfig:dB_3_2_original}
            The de Bruijn graph $G_{3,2}$.
        \protect\subref{subfig:dB_3_2_012002211}
            An Eulerian circuit in $G_{3,2}$ that represents the $(3,2)$-de Bruijn sequence $012002211$.
        \protect\subref{subfig:dB_3_3_012002211}
            The de Bruijn graph $G_{3,3}$. The arcs in the Hamiltonian cycle of the sequence $012002211$ are depicted with bold lines.
        }
    \label{fig:dB_3_2_3_3}
\end{figure}
Fig. \ref{fig:dB_3_2_3_3}\subref{subfig:dB_3_2_original}
    shows the de Bruijn graph $G_{3,2}$ used to generate the sequence.
Fig. \ref{fig:dB_3_2_3_3}\subref{subfig:dB_3_2_012002211}
    depicts the 
    Eulerian circuit on $G_{3,2}$
    that
    $s$
    corresponds to,
    which is
    $0\rightarrow 1\rightarrow 2
    \rightarrow 0 \rightarrow 0 \rightarrow 2
    \rightarrow 2 \rightarrow 1 \rightarrow 1
    \rightarrow 0$.
Fig. \ref{fig:dB_3_2_3_3}\subref{subfig:dB_3_3_012002211}
    illustrates the de Bruijn graph 
    $G_{3,3}$,
    where
    $s$
    corresponds to the Hamiltonian cycle
    $01\rightarrow 12\rightarrow 20 \rightarrow 00
    \rightarrow 02 \rightarrow 22 \rightarrow 21
    \rightarrow 11 \rightarrow 10 \rightarrow 01$.
The length-$3$ substring $012$ in $s$
    corresponds to the two-step walk
    $0\rightarrow 1\rightarrow 2$ in the Eulerian cycle on $G_{3,2}$
    and the arc $01\rightarrow 12$
    in the Hamiltonian cycle on $G_{3,3}$.
\end{example}

We also review the balanced de Bruijn sequence
    studied in \cite{Balanced_dB_Etzion}.
\begin{definition}\label{def:bal_deB}
    \cite[Definition 4]{Balanced_dB_Etzion}
    Let $k$ and $\mathcal{A}$ be as before and let $b\geq 1$.
    A circular sequence $\mathbf{s}=(s_0,s_1,\ldots,s_{b\sigma^k-1})$ is called
        a \emph{$b$-balanced $(\sigma,k)$-de Bruijn sequence}
        if each sequence in
        $\mathcal{A}^k$
        appears as a
        circular
        substring of $\mathbf{s}$ exactly $b$ times.
\end{definition}

\section{Generalized Orthogonal de Bruijn Sequences}

Our first result pertains to a generalization of orthogonal de Bruijn sequences, defined below.
\begin{definition}\label{def:ell_ortho}
    Let $\ell\geq 1$.
    A collection of $(\sigma,k)$-de Bruijn sequences $\mathcal{C}=\{\mathbf{s}_1,\ldots,\mathbf{s}_M\}$
    is called
    \emph{$\ell$-orthogonal}
    if each length-$(k+1)$ string in $\mathcal{A}^{k+1}$
    appears at most $\ell$ times in $\mathcal{C}$.
\end{definition}

Let $\Omega_\ell(\sigma,k)$ denote the maximum cardinality of 
    a collection of $\ell$-orthogonal $(\sigma,k)$-de Bruijn sequences.
We have the following bounds of
   $\Omega_\ell(\sigma,k)$:
\begin{proposition}\label{prop:ell_ortho_ub}
    One has $\Omega_{\ell}(\sigma,k)\leq \ell(\sigma-1)$.
\end{proposition}

\begin{IEEEproof}
The proof follows a similar argument as in~\cite[Corollary 4]{Ortho_dB_Skiena}. A collection of $\ell$-orthogonal $(\sigma,k)$-de Bruijn sequence
    corresponds to
    a collection of
    Hamiltonian cycles in $G_{\sigma,k+1}$
    s.t. 
    each arc is used at most $\ell$ times.
Note that
    the all-zero vertex $0^k$ in $G_{\sigma,k+1}$
    (assuming $\mathcal{A}=[0,\sigma-1])$
    has
    $\sigma-1$ inward arcs,
    $\sigma-1$ outward arcs,
    and a loop.
Also note that a Hamiltonian cycle cannot involve a loop.
Therefore,
    by the pigeonhole principle,
    any collection of more than $\ell(\sigma-1)$ 
    Hamiltonian cycles on $G_{\sigma,k+1}$
    uses at least one of the $\sigma-1$ inward arcs
    of $0^k$ more than $\ell$ times.
\end{IEEEproof}

\begin{theorem}\label{thm:ell_ortho_lb}
If $\ell\leq\sigma^{k-1}$
    and $\sigma\geq 3$,
    then
   $$\Omega_{\ell}(\sigma,k)\geq 2\ell; \;\;\;
        \Omega_{\ell}(\sigma,k)\geq \ell\lfloor \sigma/2 \rfloor.$$
        \end{theorem}
        
This result is intuitively expected, despite its proof being nontrivial. Before proceeding with the proof,
    we introduce the concept of ``wiring'' described in ~\cite{Ortho_dB_Skiena} and ~\cite{Compatible_Euler_Fleischner}.
\begin{definition}\label{def:wiring}
Let $G$ be a direct graph with an  Eulerian circuit $C$,
    and let $v$ be a vertex in $G$. For a graph to be Eulerian, each vertex must have the same in-degree and out-degree, so that in-arc and out-arc are paired up in the circuit.
A \emph{wiring} of $v$ \cite{Ortho_dB_Skiena},
    or a \emph{transition system}
    at $v$ \cite{Compatible_Euler_Fleischner},
    is a $1$-regular bipartite undirected graph (i.e., a matching) 
    between two vertex sets representing the in-arcs and out-arcs of $v$.
More precisely, the wiring of $v$ induced by $C$,
    denoted by $W(v,C)$,
    is a wiring such that
    an in-arc $a_i$
    connects to an out-arc $a_o$
    if and only if 
    the Eulerian circuit $C$
    uses $a_i$ and $a_o$ in consecutive order. In the case that $v$ has a self-loop,
    we treat that loop as both an in-arc and an out-arc.
\end{definition}

We say two Eulerian circuits are \emph{compatible}~\cite{Compatible_Euler_Fleischner}
    if the induced wirings are edge-disjoint at each vertex, i.e., two Eulerian circuits are compatible 
    if they do not use the same in-out arc pair
    at any vertex. This leads to the following characterization
    of orthogonal de Bruijn sequences:
A collection of $(\sigma,k)$-de Bruijn sequences is orthogonal
    iff their corresponding  Eulerian circuits in $G_{\sigma,k}$
    are pairwise compatible.

In our proof of Theorem~\ref{thm:ell_ortho_lb}, we will make use of the following lemmas regarding wirings.
\begin{lemma}\label{lem:rewiring}
    \cite[Theorem 7]{Ortho_dB_Skiena}
    Let $G=(V,A)$ be a directed graph with
        Eulerian circuits $C$
        and let $v\in V$.
    Let
        $\textnormal{deg}(v)$
        denote the in-out-degree of $v$
        (i.e. the in-degree of $v$,
        which is equal to the out-degree of $v$).
    If
        $\textnormal{deg}(v)\geq 3$,
        there exists an Eulerian circuit $C'$, denoted by $C'(v,C),$
        such that
    \begin{itemize}
        \item
            $W(v,C')$
            and
            $W(v,C)$
            are edge-disjoint;
        \item 
            $W(v',C')=W(v',C)$ for all $v'\in V\setminus\{v\}$.
    \end{itemize}
    In words,
        one can rewire $v$
        with respect to $C$
        to obtain $C'$
        so that
        $C'$ does not
        use the same
        in-out arc pair of $v$
        as $C$;
        the wirings at other vertices in $C'$ remain the same as in $C$.
\end{lemma}

\begin{lemma}\label{lem:rewiring_general}
    \cite[Theorem 8]{Ortho_dB_Skiena}
    \cite[Theorem 1]{Compatible_Euler_Fleischner}
    Let $G=(V,A)$ be an directed graph with an Eulerian circuit $C$,
        and let $C_1,C_2,\ldots,C_t$
        be
        compatible Eulerian circuits
        of $G$. Furthermore, let $v\in V$.
    If $t\leq \lfloor \text{deg}(v)/2 \rfloor-1 $,
        then
        there exists an Eulerian circuit $C'$, denoted by $C'(v,C|C_1,\ldots,C_t)$
        such that
    \begin{itemize}
        \item
            $W(v,C')$
            and
            $W(v,C_i)$          
            are edge-disjoint
            for $1\leq i \leq t$;
        \item 
            $W(v',C')=W(v',C)$ for all $v'\in V\setminus\{v\}$.
    \end{itemize}
    Intuitively speaking,
        given the current Eulerian circuit $C$
        and previous circuits $C_1,\ldots,C_t$,
        one can rewire $v$
        with respect to $C$
        to get $C'$
        such that
        none of $C_1,\ldots,C_t$
        uses the same
        in-out arc pair of $v$
        as $C'$.
\end{lemma}

\begin{IEEEproof}[Proof of Theorem \ref{thm:ell_ortho_lb}]

We first prove $\Omega_{\ell}(\sigma,k) \geq \ell\lfloor\sigma/2\rfloor$
    for $\sigma \geq 4$.
Let $C_{1,1}$ be an Eulerian circuit of $G_{\sigma,k}$.
We arbitrarily
    partition the vertices in $G_{\sigma,k}$
    into $\ell$ groups, say  $\mathcal{V}_1,\ldots,\mathcal{V}_{\ell}$.
Then,
    for $j \in [2,\ell],$ we
    recursively define
\begin{align}
    C_{1,j} \coloneqq C'(\mathcal{V}_{j-1},C_{1,j-1}|C_{1,1}),
    \label{eq:C_1_j}
\end{align}
where for a collection of vertices $\mathcal{V}$
the notation $C'(\mathcal{V},C|C_1,\ldots,C_t)$ refers to 
an Eulerian circuit
obtained by
rewiring each vertex in $\mathcal{V}$ with respect to $C$
given $C_1,\ldots,C_t$.
Writing $\mathcal{V}=\{v_1,\ldots,v_q\}$, we define $C'(\mathcal{V},C|C_1,\ldots,C_t)\coloneqq C^{(q)}$,
    where
    $C^{(0)}\coloneqq C$
    and
    $C^{(q')}\coloneqq C'(v_{q'},C^{(q'-1)}|C_1,\ldots,C_t),$
    for $q'\in [1,q]$.
\begin{remark}
    It is noteworthy that
        different labeling of elements in $\mathcal{V}$
        may lead to different
        $C'(\mathcal{V},C|C_1,\ldots,C_t)$.
    That is,
        $C'(\mathcal{V},C|C_1,\ldots,C_t)$
        is not unique.
    However,
        in this proof,
        we can use any one of $C'(\mathcal{V},C|C_1,\ldots,C_t)$
        to construct the circuits of interest.
\end{remark}
Next,
    denote $K\coloneqq \lfloor\sigma/2\rfloor$.
Note that
    every vertex in $G_{\sigma,k}$
    has in-out-degree $\sigma$.
Thus,
    by
    Lemma 
    \ref{lem:rewiring_general},
    for any Eulerian circuit $C$
    and any vertex $v$
    of $G_{\sigma,k}$,
    one can 
    always rewire $v$
    with respect to
    $C$
    given
    any colleciton of
    less than $K$
    compatible Eulerian circuits.
This
    ensures
    that
    the following recursive definitions hold: For $i\in [2,K-1],$
\begin{align}
    C_{i,1}
    \coloneqq
    C'(\mathcal{V}_{\ell},C_{i-1,\ell}|C_{1,1},\ldots,C_{i-1,1}),
\end{align}
    and for $j\in [2,\ell],$ 
\begin{align}
    C_{i,j}
    \coloneqq
    C'(\mathcal{V}_{j-1},C_{i,j-1}|C_{1,1},\ldots,C_{i,1}).
\end{align}
Next,
    define
    \begin{align}
    C_{K,1}
    \coloneqq
    C'(\mathcal{V}_{\ell},C_{K-1,\ell}|C_{1,1},\ldots,C_{K-1,1}),
    \label{eq:def_C_K_1}
\end{align}
    and for $j\in [2,\ell],$ let
\begin{align}
    C_{K,j}
    \coloneqq
    C'(\mathcal{V}_{j-1},C_{K,j-1}|C_{2,1},\ldots,C_{K,1}).
    \label{eq:def_C_K_j}
\end{align}
Note that in~\eqref{eq:def_C_K_j}
    the conditioned circuits
    start from $C_{2,1}$
    instead of $C_{1,1}$. We claim that the following collection $\mathcal{C}\coloneqq 
    \{
        C_{i,j}
    ~|~
        i\in [1,K], j\in [1,\ell] \}$ 
    is $\ell$-orthogonal. 
Let $v$ be in the $m$th group of vertices, $\mathcal{V}_m,$
    for some $m\in [1,\ell]$.
Observe that
\begin{itemize}
    \item 
        For each $i\in [1,K-1]$,
            the wiring of 
            $v$
            is the same
            in all $C_{i,m+1},\ldots,C_{i,\ell},C_{i+1,1},\ldots,C_{i+1,m}$.
        Denote it by $W_i$. Denote by $W_0$ and $W_{K}$ the shared wiring of $v$ in $C_{1,1},\ldots,C_{1,m}$
        and  $C_{K,m+1},\ldots,C_{K,\ell}$,
        respectively.
    \item
        The wirings $W_i$ and $W_{i'}$ are edge-disjoint for $i\in [0,K]$, 
        $i'\in [1,K-1]$
        and $i\neq i'$.
        Thus,
            each in-out pair of $v$
            used in any  $W_1,\ldots, W_{\ell-1}$
            is used
            exactly $\ell$ times in $\mathcal{C}$.
    \item 
        Even if $W_0$ and $W_K$ share some edge in their bipartite graphs,
            that edge is used only $\ell$ times
            ($m$ times by $W_0$
            and $\ell-m$ times
            by $W_K$).
        This
            establishes
            the claim.
\end{itemize}
 
Now we prove $\Omega_{\ell}(\sigma,k)\geq 2\ell$ as long as $\sigma \geq 3$.
Again,
    let $C_{1,1}$ be an Eulerian circuit of $G_{\sigma,k}$.
We recursively apply Lemma \ref{lem:rewiring}
    to define for $j\in[2,\ell]$ that
\begin{align}
    C_{1,j} \coloneqq C'(\mathcal{V}_{j-1},C_{1,j-1}),
    \label{eq:C_1_j_sigma_3}
\end{align}
Next,
    define
    \begin{align}
    C_{2,1}
    \coloneqq
    C'(\mathcal{V}_{\ell},C_{1,\ell}),
    \label{eq:def_C_K_1_sigma_3}
\end{align}
    and for $j\in [2,\ell],$ let
\begin{align}
    C_{2,j}
    \coloneqq
    C'(\mathcal{V}_{j-1},C_{2,j-1}).
    \label{eq:def_C_K_j_sigma_3}
\end{align}
A similar argument shows that
    $\mathcal{C}\coloneqq \{C_{i,j}~|~i\in\{1,2\},j\in[1,\ell]\}$
    is $\ell$-orthogonal.
\end{IEEEproof}

\begin{example}
This example demonstrates the rewiring process from  the proof of Theorem~\ref{thm:ell_ortho_lb}.
Consider $k=2$, $\mathcal{A}=\{0,1,2\}$,
    and $\ell=2$.
We seek to find $2\ell=4$ Eulerian circuits on $G_{3,2}$
    such that
    each in-out arc pair of each vertex
    is used at most twice.
We first partition $V(G_{3,2})=\{0,1,2\}$
    into 
    $\mathcal{V}_1=\{0\}$
    and
    $\mathcal{V}_2=\{1,2\}$.
Also, we select the starting circuit
    $C_{1,1}$
    to be the Eulerian circuit in $G_{3,2}$ for the $(3,2)$-de Bruijn sequence $012002211$. With a slight abuse of notation we write $C_{1,1}=012002211$.
According to~\eqref{eq:C_1_j},
    ~\eqref{eq:def_C_K_1},
    and
    ~\eqref{eq:def_C_K_j},
    the rewired circuits are 
    $C_{1,2}=012022110$, $C_{2,1}=011220210$,
    and
    $C_{2,2}=011220021$ (Fig.~\ref{fig:rewiring}). Comparing Fig.~\ref{fig:dB_3_2_3_3}\subref{subfig:dB_3_2_012002211} and Fig.~\ref{fig:rewiring},
    we see that
    each vertex has
    two edge-disjoint wirings
    (gray and violet),
    each of which appears twice in the collection
    $\{C_{1,1},C_{1,2},C_{2,1},C_{2,2}\}$.
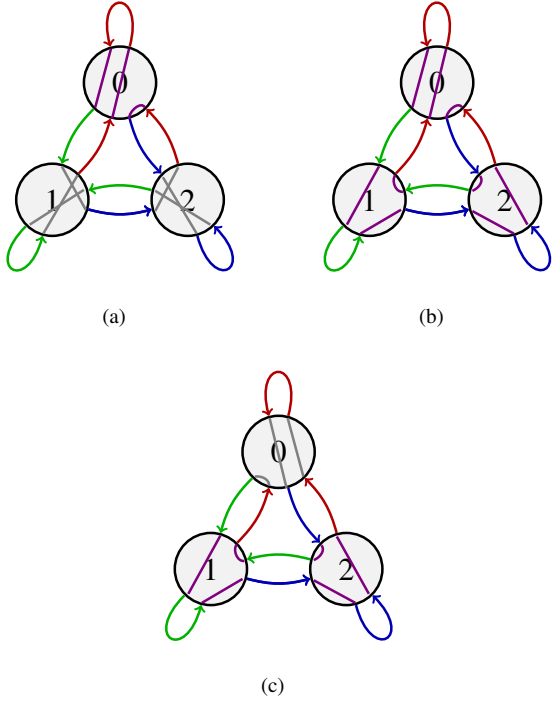
\begin{figure}
    \centering
    \subfloat[][]{
        \scalebox{0.6}{
            \ifdefined\DrawTikz
                \newcommand{\LD}{\langle}
\newcommand{\RD}{\rangle}
\begin{tikzpicture}[scale=1]
\node[draw, circle, color=black, fill=black!5, ultra thick,minimum size=1.6cm] (0) at (0,1.5*1.732) {\huge 0};
\node[draw, circle, color=black, fill=black!5, ultra thick,minimum size=1.6cm] (1) at (-1.5,0) {\huge 1};
\node[draw, circle, color=black, fill=black!5, ultra thick,minimum size=1.6cm] (2) at (1.5,0) {\huge 2};

\draw[->, ultra thick, color = green!70!black] (0) to [out=225,in=75] (1);
\draw[->, ultra thick, color = blue!70!black] (1) to [out=-15,in=195] (2);
\draw[->, ultra thick, color = blue!70!black] (1) to [out=-15,in=195] (2);
\draw[->, ultra thick, color = red!70!black] (2) to [out=105,in=-45] (0);

\draw[->, ultra thick, color = blue!70!black] (0) to [out=-75,in=135] (2);
\draw[->, ultra thick, color = green!70!black] (2) to [out=165,in=15] (1);
\draw[->, ultra thick, color = red!70!black] (1) to [out=45,in=255] (0);

\draw[->, ultra thick, color = red!70!black] (0) to [out=75,in=105,loop] (0);
\draw[->, ultra thick, color = green!70!black] (1) to [out=225,in=255,loop] (1);
\draw[->, ultra thick, color = blue!70!black] (2) to [out=285,in=315,loop] (2);







\draw[gray, ultra thick] (-1.29,0.74) -- (-0.76,-0.22);
\draw[gray, ultra thick] (0.78,-0.25) -- (1.29,0.72);
\draw[gray, ultra thick] (0.95,0.5) -- (1.7,-0.75);
\draw[gray, ultra thick] (2,-0.55) -- (0.78,0.2);
\draw[gray, ultra thick] (-0.8,0.22) -- (-2,-0.55);
\draw[gray, ultra thick] (-1.7,-0.75) -- (-0.9375,0.55);


\draw[violet, ultra thick] (0.55,2.05) to [out=120,in=90] (0.2,1.8);
\draw[violet, ultra thick] (-0.175,1.8) -- (0.2,3.35);
\draw[violet, ultra thick] (-0.2,3.35) -- (-0.55,2.05);

\end{tikzpicture}%
            \fi
        }
        \label{subfig:rewiring_C_1_2}
    }
    \subfloat[][]{
        \scalebox{0.6}{
            \ifdefined\DrawTikz
                \newcommand{\LD}{\langle}
\newcommand{\RD}{\rangle}
\begin{tikzpicture}[scale=1]
\node[draw, circle, color=black, fill=black!5, ultra thick,minimum size=1.6cm] (0) at (0,1.5*1.732) {\huge 0};
\node[draw, circle, color=black, fill=black!5, ultra thick,minimum size=1.6cm] (1) at (-1.5,0) {\huge 1};
\node[draw, circle, color=black, fill=black!5, ultra thick,minimum size=1.6cm] (2) at (1.5,0) {\huge 2};

\draw[->, ultra thick, color = green!70!black] (0) to [out=225,in=75] (1);
\draw[->, ultra thick, color = blue!70!black] (1) to [out=-15,in=195] (2);
\draw[->, ultra thick, color = blue!70!black] (1) to [out=-15,in=195] (2);
\draw[->, ultra thick, color = red!70!black] (2) to [out=105,in=-45] (0);

\draw[->, ultra thick, color = blue!70!black] (0) to [out=-75,in=135] (2);
\draw[->, ultra thick, color = green!70!black] (2) to [out=165,in=15] (1);
\draw[->, ultra thick, color = red!70!black] (1) to [out=45,in=255] (0);

\draw[->, ultra thick, color = red!70!black] (0) to [out=75,in=105,loop] (0);
\draw[->, ultra thick, color = green!70!black] (1) to [out=225,in=255,loop] (1);
\draw[->, ultra thick, color = blue!70!black] (2) to [out=285,in=315,loop] (2);







\draw[violet, ultra thick] (0.55,2.05) to [out=120,in=90] (0.2,1.8);
\draw[violet, ultra thick] (-0.175,1.8) -- (0.2,3.35);
\draw[violet, ultra thick] (-0.2,3.35) -- (-0.55,2.05);

\draw[violet, ultra thick] (-1.7,-0.75) -- (-0.8,-0.22);
\draw[violet, ultra thick] (-1.29,0.74) -- (-2,-0.55);
\draw[violet, ultra thick] (-0.76,0.18) to [out=190, in=230] (-0.925,0.55);



\draw[violet, ultra thick] (0.78,-0.25) -- (1.7,-0.75);
\draw[violet, ultra thick] (0.95,0.525) to [out=310, in=20] (0.78,0.2);
\draw[violet, ultra thick] (2,-0.55) -- (1.29,0.72);

\end{tikzpicture}%
            \fi
        }
        \label{subfig:rewiring_C_2_1}
    }\\
        \subfloat[][]{
        \scalebox{0.6}{
            \ifdefined\DrawTikz
                \newcommand{\LD}{\langle}
\newcommand{\RD}{\rangle}
\begin{tikzpicture}[scale=1]
\node[draw, circle, color=black, fill=black!5, ultra thick,minimum size=1.6cm] (0) at (0,1.5*1.732) {\huge 0};
\node[draw, circle, color=black, fill=black!5, ultra thick,minimum size=1.6cm] (1) at (-1.5,0) {\huge 1};
\node[draw, circle, color=black, fill=black!5, ultra thick,minimum size=1.6cm] (2) at (1.5,0) {\huge 2};

\draw[->, ultra thick, color = green!70!black] (0) to [out=225,in=75] (1);
\draw[->, ultra thick, color = blue!70!black] (1) to [out=-15,in=195] (2);
\draw[->, ultra thick, color = blue!70!black] (1) to [out=-15,in=195] (2);
\draw[->, ultra thick, color = red!70!black] (2) to [out=105,in=-45] (0);

\draw[->, ultra thick, color = blue!70!black] (0) to [out=-75,in=135] (2);
\draw[->, ultra thick, color = green!70!black] (2) to [out=165,in=15] (1);
\draw[->, ultra thick, color = red!70!black] (1) to [out=45,in=255] (0);

\draw[->, ultra thick, color = red!70!black] (0) to [out=75,in=105,loop] (0);
\draw[->, ultra thick, color = green!70!black] (1) to [out=225,in=255,loop] (1);
\draw[->, ultra thick, color = blue!70!black] (2) to [out=285,in=315,loop] (2);






\draw[violet, ultra thick] (-1.7,-0.75) -- (-0.8,-0.22);
\draw[violet, ultra thick] (-1.29,0.74) -- (-2,-0.55);
\draw[violet, ultra thick] (-0.76,0.18) to [out=190, in=230] (-0.925,0.55);



\draw[violet, ultra thick] (0.78,-0.25) -- (1.7,-0.75);
\draw[violet, ultra thick] (0.95,0.525) to [out=310, in=20] (0.78,0.2);
\draw[violet, ultra thick] (2,-0.55) -- (1.29,0.72);


\draw[gray, ultra thick] (-0.21,1.85) to[out=90,in=7.5] (-0.525,2.05);

\draw[gray, ultra thick] (0.55,2.05) -- (0.2,3.35);
\draw[gray, ultra thick] (-0.2,3.35) -- (0.175,1.8);

\end{tikzpicture}%
            \fi
        }
        \label{subfig:rewiring_C_2_2}
    }
    \caption{
        \protect\subref{subfig:rewiring_C_1_2}
            The circuit $C_{1,2}=012022110$ obtained by
                rewiring $C_{1,1}=012002211$
                at the vertex $0$, 
                given $C_{1,1}$.
        \protect\subref{subfig:rewiring_C_2_1}
            The circuit $C_{2,1}=011220210$
                obtained by
                rewiring $C_{1,2}$
                at the vertices $1$ and $2$, 
                given $C_{1,1}$.
        \protect\subref{subfig:rewiring_C_2_2}
            The circuit $C_{2,2}=011220021$
                obtained by
                rewiring $C_{2,1}$
                at the vertex $0$, 
                given $C_{2,1}$.
        }
    \label{fig:rewiring}
\end{figure}
\end{example}

\section{Orthogonal Balanced de Bruijn Sequences}\label{sec:ortho_bal_deB}

We start with several
    definitions
    relevant to
    the balanced de Bruijn sequence
    defined in Definition \ref{def:bal_deB}.

\begin{definition}\label{def:ortho_bal}
 A $b$-balanced $(\sigma,k)$-de Bruijn sequence $\mathbf{s}$
        is \emph{self-orthogonal}
        if any length-$(k+1)$ string in $\mathcal{A}$
        appears at most once in $\mathbf{s}$.
A collection of $b$-balanced $(\sigma,k)$-de Bruijn sequences $\mathcal{C}=\{\mathbf{s}_1,\ldots,\mathbf{s}_M\}$ is called 
    \emph{orthogonal}
    if each length-$(k+1)$ string in $\mathcal{A}^{k+1}$
    appears at most once in $\mathcal{C}$. Clearly, a $b$-balanced de Bruijn sequence
        in an orthogonal collection is self-orthogonal.
\end{definition}

\begin{definition}\label{def:b_ckt}
    Let $b\geq 1$, $G$ be a directed graph,
        and let $C$ be a circuit of $G$.
    We say that
        $C$ is a \emph{$b$-circuit} of $G$
        if $C$ visit each vertex of $G$ exactly $b$ times. A $b$-circuit is sometimes referred to as an exact $b$-walk~\cite{k_walks_jackson}.
\end{definition}

\begin{proposition}\label{prop:bal_deB_b_ckt}
    There is a 1-1  correspondence between
        self-orthogonal $b$-balanced $(\sigma,k)$-de Bruijn sequences
        and
        $b$-circuits on $G_{\sigma,k+1}$.
    A collection of $b$-balanced $(\sigma,k)$-de Bruijn sequences
        is orthogonal
        iff
        each sequence is self-orthogonal
        and
        their corresponding $b$-circuits are arc-disjoint.
\end{proposition}
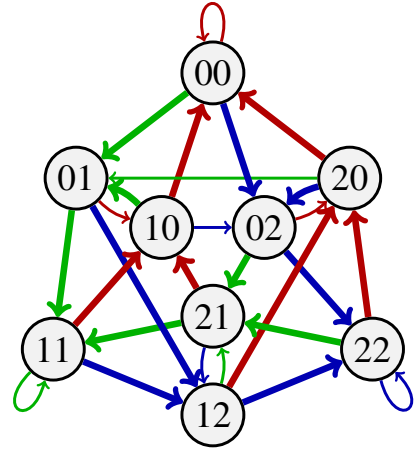
\begin{figure}
   \centering
   \scalebox{0.7}{
       \ifdefined\DrawTikz
           \begin{tikzpicture}[scale=1]
\node[draw, circle, color=black, fill=black!5, ultra thick,minimum size=1cm] (00) at (90:3.5cm) {\huge 00};
\node[draw, circle, color=black, fill=black!5, ultra thick,minimum size=1cm] (11) at (210:3.5cm) {\huge 11};
\node[draw, circle, color=black, fill=black!5, ultra thick,minimum size=1cm] (22) at (330:3.5cm) {\huge 22};
\node[draw, circle, color=black, fill=black!5, ultra thick,minimum size=1cm] (01) at (150:3cm) {\huge 01};
\node[draw, circle, color=black, fill=black!5, ultra thick,minimum size=1cm] (12) at (270:3cm) {\huge 12};
\node[draw, circle, color=black, fill=black!5, ultra thick,minimum size=1cm] (20) at (30:3cm) {\huge 20};
\node[draw, circle, color=black, fill=black!5, ultra thick,minimum size=1cm] (10) at (150:1.125cm) {\huge 10};
\node[draw, circle, color=black, fill=black!5, ultra thick,minimum size=1cm] (21) at (270:1.125cm) {\huge 21};
\node[draw, circle, color=black, fill=black!5, ultra thick,minimum size=1cm] (02) at (30:1.125cm) {\huge 02};


\draw[->, ultra thick, color = red!70!black] (00) to [out=75, in=105, loop] (00);
\draw[->, line width = 0.125cm, color = red!70!black] (10) -- (00);
\draw[->, line width = 0.125cm, color = red!70!black] (20) -- (00);

\draw[->, line width = 0.125cm, color = green!70!black] (01) -- (11);
\draw[->, ultra thick, color = green!70!black] (11) to[out=225, in=255, loop] (11);
\draw[->, line width = 0.125cm, color = green!70!black] (21) -- (11);

\draw[->, line width = 0.125cm, color = blue!70!black] (02) -- (22);
\draw[->, line width = 0.125cm, color = blue!70!black] (12) -- (22);
\draw[->, ultra thick, color = blue!70!black] (22) to[out=285, in=315, loop] (22);

\draw[->, line width = 0.125cm, color = green!70!black] (00) -- (01);
\draw[->, line width = 0.125cm, color = green!70!black] (10) to[out=135,in=-15] (01);
\draw[->, ultra thick, color = green!70!black] (20) -- (01);

\draw[->, line width = 0.125cm, color = blue!70!black] (01) -- (12);
\draw[->, line width = 0.125cm, color = blue!70!black] (11) -- (12);
\draw[->, ultra thick, color = blue!70!black] (21) to[out=255,in=105]  (12);

\draw[->, ultra thick, color = red!70!black] (02) to[out=15,in=225] (20);
\draw[->, line width = 0.125cm, color = red!70!black] (12) -- (20);
\draw[->, line width = 0.125cm, color = red!70!black] (22) -- (20);

\draw[->, line width = 0.125cm, color = blue!70!black] (00) -- (02);
\draw[->, ultra thick, color = blue!70!black] (10) -- (02);
\draw[->, line width = 0.125cm, color = blue!70!black] (20) to[out=195,in=45] (02);

\draw[->, line width = 0.125cm, color = green!70!black] (02) -- (21);
\draw[->, ultra thick, color = green!70!black] (12) to[out=75,in=285] (21);
\draw[->, line width = 0.125cm, color = green!70!black] (22) -- (21);

\draw[->, ultra thick, color = red!70!black] (01) to[out=-45,in=165] (10);
\draw[->, line width = 0.125cm, color = red!70!black] (11) -- (10);
\draw[->, line width = 0.125cm, color = red!70!black] (21) -- (10);





\end{tikzpicture}%
       \fi
   } 
   \caption{
       The de Bruijn graph $G_{3,3}$. The bold arcs correspond to the $2$-circuit
           representing
           the $2$-balanced
           $(3,2)$-de Bruijn
           sequence
           $002211012001122021$.
       }
       \vspace{-0.15in}
   \label{fig:bdB_2_3_3}
\end{figure}
\begin{example}
Consider $k=2$, $\mathcal{A}=\{0,1,2\}$,
    and
    $b=2$.
The circular sequences $\mathbf{s}_1=000111222020212101$ 
    and $\mathbf{s}_2=002211012001122021$ are
    two 
    $2$-balanced $(3,2)$-de Bruijn sequences
    since each length-$2$ sequence in $\{0,1,2\}$
    appears exactly twice in both $\mathbf{s}_1$ and $\mathbf{s}_2$.
The sequence $\mathbf{s}_1$ is not self-orthogonal
    since the length-$3$ sequence $202$
    appears twice in $\mathbf{s}_1$.
On the other hand,
    $\mathbf{s}_2$ is self-orthogonal,
    and it
    corresponds to the $2$-circuit
    $00\rightarrow 02\rightarrow 22\rightarrow 21
            \rightarrow 11\rightarrow 10\rightarrow 01
            \rightarrow 12\rightarrow 20\rightarrow 00
            \rightarrow 01\rightarrow 11\rightarrow 12
            \rightarrow 22\rightarrow 20\rightarrow 02
            \rightarrow 21\rightarrow 10\rightarrow 00$
    on $G_{3,3}$
    (see Fig.~\ref{fig:bdB_2_3_3}).
\end{example}

Fix $\sigma$.
Observe that that when $b$ increases, the number of length-($k+1$) sequences
    in any two $b$-balanced $(\sigma,k)$-de Bruijn sequences increases as well,
    making them less likely to be orthogonal.
This motivates the following definition:
For $c\geq 2$,
    define 
    $\Sigma(c,b,k)$ to be
    the minimum $\sigma$
    such that there exist $c$ orthogonal
    $b$-balanced $(\sigma,k)$-de Brujin sequences. 

We start by establishing a lower bound of $\Sigma(b,k)$.

\begin{proposition}\label{prop:Sigma_lb}
    If $b\geq 2$, then $\Sigma(c,b,k)\geq cb$.
\end{proposition}
\begin{IEEEproof}
Similarly to the proof of Proposition~\ref{prop:ell_ortho_ub},
    we observe that
    there are 
    $\sigma$ outgoing arcs
    for each vertex in $G_{\sigma,k+1}$.
By Proposition~\ref{prop:bal_deB_b_ckt},
    a self-orthogonal $b$-balanced de Bruijn sequence
    uses exactly $b$ ingoing and outgoing arcs at each vertex.
Therefore,
    since $c$ orthogonal $b$-balanced de Bruijn sequences
    share no arcs,
    they must use $cb$ distinct
    outgoing arcs at each vertex.
This is impossible if $\sigma<cb$.
\end{IEEEproof}

We also have the following upper bound on $\Sigma(c,b,k).$ 

\begin{theorem}\label{thm:Sigma_ub}
For $b\geq 2$, $\Sigma(c,b,k)$ is at most the smallest prime power that is greater than or equal to $cb$.
Furthermore,
    $\Sigma(c,b,k)=cb$
    if each prime factor of $c$ divides $b$.
\end{theorem}

To prove Theorem \ref{thm:Sigma_ub},
    we establish the following lemmas.
\begin{lemma}\label{lem:ortho_bal_power_prime}
Assume
    $b\geq 2$,
    $\sigma=p^m$ for some prime $p$
    and positive integer $m$,
    and $\sigma\geq cb$.
Then there exist
    $c$ orthogonal $b$-balanced
    $(\sigma,k)$-de Bruijn sequences.
\end{lemma}
\begin{IEEEproof}
    Since $\sigma$ is a prime power, by \cite[Lemma 2]{Arc_Disjoint_Rowley_Bose},
        there exist
        cycles
        $C_0,\ldots,C_{\sigma-1}$
        on $G_{\sigma,k+1}$
        such that:
        \begin{itemize}
            \item
                For
                    $i,j\in [0,\sigma-1]$
                    with $i\neq j$,
                    the cycles
                    $C_i$ and $C_j$ are 
                    arc-disjoint.
            \item
                For
                    $i\in [0,\sigma-1]$,
                    the cycle $C_i$
                    does not visit
                    the vertex
                    representing the sequence $ii\cdots i$
                    (assuming $\mathcal{A}=\{0,\ldots,\sigma-1\}$)
                    and visits
                    all other vertices
                    exactly once.
        \end{itemize}
    Then,
        for each $\tau\in [0,c-1],$
        we can construct a circuit $\hat{C}_{\tau}$ as follows. Let  
        $\tilde{C}_{b\tau+i}$ be the cycle $C_{b\tau+i}$
        with an additional transition 
        following the loop on the vertex
        representing the all-$(b\tau+i+1)$ sequence,
        where $i=0,\ldots,b-2$.
    Note that this definition is valid since
        $C_{b\tau+i}$ must visit the all-$(b\tau+i+1)$ sequence.
    Also, let  
        $\tilde{C}_{b\tau+b-1}$ be the cycle $C_{b\tau+b-1}$
        with an additional transition
        following the loop at  the all-$(b\tau)$ vertex.
    Then, 
        let $\hat{C}_{\tau}$ be a combination of 
        $\tilde{C}_{b\tau},\ldots,\tilde{C}_{b\tau+b-1}$ (a combination of
            $\tilde{C}_{b\tau},\ldots,\tilde{C}_{b\tau+b-1}$ is 
            a closed walk whose arcs
            are exactly those in
            $\tilde{C}_{b\tau},\ldots,\tilde{C}_{b\tau+b-1}$).
        Since
            each pair of $\tilde{C}_i,\tilde{C}_j$
            meets at some vertex,
            such a closed walk always exists. Then,  observe that a) the arcs used in $\hat{C}_{\tau}$
                are those in $C_{b\tau},\ldots,C_{b\tau+b-1}$
                and the loops at the all-$(b\tau+i)$ vertices
                for $i\in [0,b-1]$; b)
$C_{b\tau},\ldots,C_{b\tau+b-1}$
                are arc-disjoint
                and do not use any loops; c) for $i\in [0,b-1]$,
                the all-$(b\tau+i)$ vertex
                is visited $b$ times in $\hat{C}_{\tau}$ (twice by $\hat{C}_{b\tau+((i-1)\mod b)}$,
                never by $\hat{C}_{b\tau+i}$,
                and once by each remaining $\hat{C}_{b\tau+j}$); d) each of the remaining vertices in $G_{\sigma,k+1}$
                is also visited $b$ times in $\hat{C}_{\tau}$ (once by each $\hat{C}_{b\tau+j}$). Therefore,
        $\hat{C}_{\tau}$ is a $b$-circuit on $G_{\sigma,k+1}$.
    The arc-disjointness and no-loop property
        of the $C_i$s imply that
        $\hat{C}_{\tau}$ and $\hat{C}_{\tau'}$
        are arc-disjoint
        for $\tau\neq\tau'$.
    Thus, $\hat{C}_0,\ldots,\hat{C}_{c-1}$
        are orthogonal $b$-balanced
        $(\sigma,k)$-de Bruijn sequences.
\end{IEEEproof}
\begin{lemma}\label{lem:b1b2_ckt}
Let $G_1=(V_1,A_1)$
    and $G_2=(V_2,A_2)$
    be two directed graphs.
Write $N_1\coloneqq |V_1|$ and $N_2\coloneqq |V_2|$. 
Assume 
    $G_1$ has
    $n_1$ arc-disjoint $b_1$-circuits
    and
    $G_2$ has
    $n_2$ arc-disjoint $b_2$-circuits.
If $N_1b_1$ and $N_2b_2$ are coprime,
    then the tensor-product graph $G_1\times G_2$
    has
    $n_1n_2$ arc-disjoint $b_1b_2$-circuits.
\end{lemma}
\begin{IEEEproof}
Let $C_1^{(1)},\ldots,C_1^{(n_1)}$ be arc-disjoint $b_1$-circuits of $G_1$
    and
    $C_2^{(1)},\ldots,C_2^{(n_2)}$ be arc-disjoint $b_2$-circuits of $G_2$.

Then, for each
    $i\in [1,n_1]$
    and
    $j\in [1,n_2]$, 
    define $C_{i,j}$ to be the subgraph of $G_1\times G_2$ with arc set
    \begin{align*}
        A_{i,j}&\coloneqq
        \left\{
            ((v_1,v_2),(v_1',v_2'))
        ~\middle|~
        \vphantom{E(C_1^{(j)}}
        \right.\nonumber\\
        &\left.
            (v_1,v_1')\in E(C_1^{(i)})
            \text{ and }
            (v_2,v_2')\in E(C_2^{(j)})
        \right\}.
    \end{align*}

We first show that $C_{i,j}$ is a circuit. 
Since $C_1^{(i)}$ is a circuit of length $N_1b_1$,
    we can traverse $C_1^{(i)}$
    following a circular sequence of distinct arcs
    $(e_1,e_2,\ldots,e_{N_1b_1})$,
    where each $e_k\in A_1$
    and
    $e_k\rightarrow e_{k+1}$.
Similarly,
    we can represent $C_2^{(j)}$
    by a circular sequence of arcs
    $(f_1,f_2,\ldots,f_{N_2b_2})$,
    where each $f_{\ell}\in A_2$
    and
    $f_{\ell}\rightarrow f_{\ell+1}$.
Then, with a slight abuse of notation, we write $A_{i,j}= \{
    (e_k,f_{\ell})
    ~|~
        k\in\llbracket 1,N_1b_1\rrbracket, 
        \ell\in\llbracket 1,N_2b_2\rrbracket
    \}.$
Now fix an arc $(e_k,f_{\ell})$ of $C_{i,j}$.
We can traverse $C_{i,j}$
    starting from  $(e_k,f_{\ell})$ and by following
    $(e_{k+1},f_{\ell+1})$,
    $(e_{k+2},f_{\ell+2})$, etc.
Then,
    the next time $(e_k,f_{\ell})$ is used 
    is after
    $\text{lcm}(N_1b_1,N_2b_2)=N_1N_2b_1b_2$ steps.
However,
    note that there are exactly $N_1N_2b_1b_2$ arcs in $C_{i,j}$.
Therefore,
    right before the next use of $(e_k,f_{\ell})$ in  $C_{i,j}$,
    each arc in $C_{i,j}$ has been traversed,
    and thus
    $C_{i,j}$ represents a single closed walk on $G_1\times G_2$.
Furthermore,
    if some arc is repeated,
    starting the traversal from that arc
    would contradict the fact that
    the next use of each arc has to happen after $N_1N_2b_1b_2$ steps. Hence, $C_{i,j}$ is a circuit. Then,
    since $C_1^{(i)}$ and $C_1^{(i')}$ are arc-disjoint for $i\neq i'$
    and
    $C_2^{(j)}$ and $C_2^{(j')}$ for $j\neq j'$,
    we have that $C_{i,j}$ and $C_{i',j'}$ are arc-disjoint
    when either $i\neq i'$ or $j\neq j'$.

To verify that each $C_{i,j}$ is a $b_1b_2$ circuit,
    it remains to show that
    each $C_{i,j}$ visits each vertex in $G_1\times G_2$
    exactly $b_1b_2$ times.
Fix a vertex $v$ in $G_1\times G_2$.
Write $v=(v_1,v_2)$ for some $v_1 \in V_1$ and $v_2\in V_2$.
On $C_1^{(i)}$,
    the vertex $v_1$
    has exactly $n_1$ out-arcs,
    say $e'_1,\ldots,e'_{b_1}$.
Similarly,
    on $C_2^{(j)}$,
    the out-arcs of vertex $v_2$ are denoted by  $f'_1,\ldots,f'_{b_2}$.
Therefore,
    the out-arcs of $(v_1,v_2)$ on $C_{i,j}$
    are of the form $(e'_{k'},f'_{\ell'})$,
    where $k'\in [1,b_1]$
    and $\ell'\in [1,b_2]$.
Since $(v_1,v_2)$ has exactly $b_1b_2$ out-edges in $C_{i,j}$,
    it is visited exactly $b_1b_2$ times in $C_{i,j}$.
Since $v$ is arbitrary,
    $C_{i,j}$ is a $b_1b_2$-circuit.

In conclusion,
   we have shown that
   $\{C_{i,j}\}_{i\in\llbracket 1,n_1 \rrbracket,j\in\llbracket 1,n_2 \rrbracket}$
   is a collection of $n_1n_2$ arc-disjoint $b_1b_2$-circuits.
\end{IEEEproof}
\begin{lemma}\label{lem:deB_iso}
The tensor-product graph
    $G_{\sigma_1,k}\times G_{\sigma_2,k}$
    is isomorphic to
    $G_{\sigma_1\sigma_2,k}$.
\end{lemma}
\begin{IEEEproof}[Sketch of the proof]
Without loss of generality,
    assume that the alphabets of
    $G_{\sigma_1,k}$, $G_{\sigma_2,k}$, and $G_{\sigma_1\sigma_2,k}$ 
    are
    $[0,\sigma_1-1]$,
    $[0,\sigma_2-1]$, and
    $[0,\sigma_1\sigma_2-1]$,
    respectively.
It is not hard to show that the following mapping is a graph isomorphism:
\begin{align}
    G_{\sigma_1\sigma_2,k} &\rightarrow G_{\sigma_1,k}\times G_{\sigma_2,k} \nonumber\\
    (s_0, \ldots, s_{k-2}) &\mapsto ((q_0,\ldots,q_{k-2}),(r_0,\ldots, r_{k-2})),
    \label{eq:isomorphism}
\end{align}
where for $i\in [0, k-2]$ we write $q_i$ and $r_i$ for the quotient and remainder of $s_i$ when divided by $\sigma_2$, respectively. Note that the construction in~\eqref{eq:isomorphism} is related to the proofs of~\cite[Lemma 3]{Arc_Disjoint_Rowley_Bose} and~\cite[Lemma 3]{Note_IJGood_Rees}.
\end{IEEEproof}

\begin{IEEEproof}[Sketch of the proof of Theorem \ref{thm:Sigma_ub}]
Lemma~\ref{lem:ortho_bal_power_prime} established the first statement in the theorem. Now, assume that each prime factor of $c$ divides $b$.
Then,
    we can write
    $c=p_1^{x_1}\cdots p_m^{x_m}$
    and
    $b=p_1^{y_1}\cdots p_m^{y_m}R$,
    where
    $p_1,\ldots,p_m$ are distinct prime numbers,
    $x_1,\ldots,x_m,y_1,\ldots,y_m$ are positive integers,
    and
    none of the $p_1,\ldots,p_m$ divides $R$.

Now consider
    the following $m+1$ de Bruijn graphs     $G_{p_1^{x_1+y_1},k+1},\ldots,G_{p_m^{x_m+y_m},k+1},$ 
    $G_{R,k+1}$.
By Lemma~\ref{lem:ortho_bal_power_prime},
    for each $i\in [1,m]$,
    $G_{p_i^{x_i+y_i},k+1}$
    has
    $p_i^{x_i}$
    arc-disjoint $p_i^{y_i}$-circuits.
We also know that
    $G_{R,k+1}$ has one $R$-circuit,
    which is essentially an Eulerian circuit on it.
Then,
    by repeatedly applying Lemma~\ref{lem:b1b2_ckt} and Lemma~\ref{lem:deB_iso},
    we can see that
    $G_{\prod_{i=1}^m p_i^{x_i+y_i}R,k+1}$
    has
    $\prod_{i=1}^m p_i^{x_i}$
    arc-disjoint
    $(\prod_{i=1}^m p_i^{y_i}R)$-circuits. Equivalently,
    $G_{cb,k+1}$
    has
    $c$
    arc-disjoint
    $b$-circuits.
Therefore,
    $\Sigma(c,b,k)\leq cb$.
We conclude that $\Sigma(c,b,k)=cb$ by invoking Proposition~\ref{prop:Sigma_lb}.
\end{IEEEproof}
\begin{example}
Consider $c=2$, $b=6$, and $k=2$ in Theorem~\ref{thm:Sigma_ub}.
In this case,
    $p_1=2$, $x_1=y_1=1$, and $R=3$. Next, examine the two de Bruijn graphs
    $G_{4,3}$ and $G_{3,3}$.
We associate $G_{3,3}$
    with the Eulerian circuit
    $E=1 0 0 0 2 0 2 1 2 2 1 0 2 2 2 0 0 1 0 1 2 1 1 2 0 1 1$.

Lemma~\ref{lem:ortho_bal_power_prime}
    implies that
    $G_{4,3}$ has two arc-disjoint $2$-circuits, constructed as follows. 
The cycles $C_0,C_1,C_2,C_3$  are
\begin{align}
    C_0 = 0 1 1 3 1 0 2 2 1 2 0 3 3 2 3,\;
    C_1 = 1 0 0 2 0 1 3 3 0 3 1 2 2 3 2,\\ \notag
    C_2 = 2 3 3 1 3 2 0 0 3 0 2 1 1 0 1,\;
    C_3 = 3 2 2 0 2 3 1 1 2 1 3 0 0 1 0.
\end{align}
Then,
    the circuit $\tilde{C}_0$
    is defined by $C_0$ following the loop on the vertex $11$.
Written as a sequence,
    $\tilde{C}_0 = 0 1 1 1 3 1 0 2 2 1 2 0 3 3 2 3$.
Note that the loop on $11$ induces the length-$3$ substring $111$ in $\tilde{C}_0$.
Similarly,
    we have
\begin{align}
    \tilde{C}_1 = 1 0 0 0 2 0 1 3 3 0 3 1 2 2 3 2,\;
    \tilde{C}_2 = 2 3 3 3 1 3 2 0 0 3 0 2 1 1 0 1,\\ \notag
    \tilde{C}_3 = 3 2 2 2 0 2 3 1 1 2 1 3 0 0 1 0.
\end{align}
Let $\hat{C}_0$ and $\hat{C}_1$
    be
    a combination of $\tilde{C}_0$ and $\tilde{C}_1$
    and
    one of $\tilde{C}_2$ and $\tilde{C}_3$,
    respectively.
For example,
    we can take
\begin{align*}
    \hat{C}_0 = 0 1 1 1 3 1 0 2 2 1 2 0 3 3 2 3 0 1 3 3 0 3 1 2 2 3 2 1 0 0 0 2,\\
    \hat{C}_1 = 2 3 3 3 1 3 2 0 0 3 0 2 1 1 0 1 2 3 1 1 2 1 3 0 0 1 0 3 2 2 2 0.
\end{align*}
Then,
    $\hat{C}_0$ and $\hat{C}_1$
    are arc-disjoin $2$-circuits on $G_{4,3}$.

Then,
    we define
    $A_0 = 3\hat{C}_0 + E$
    and
    $A_1 = 3\hat{C}_1 + E,$ 
    where we treat $\hat{C}_0$, $\hat{C}_1$ and $E$
    as infinitely long periodic sequences, 
    and with additions performed entry-wise.
The first few terms of $A_0$ are 
    $(1,3,3,3,11,\ldots)$,
    and
    the period of $A_0$ is $\text{lcm}(32,27)=864$
    since $\hat{C}_0$ has period $32$
    and $E$ has period $27$.
By Lemma \ref{lem:b1b2_ckt}
    and Lemma \ref{lem:deB_iso},
    $A_0$ and $A_1$ correspond to
    two arc-disjoint $6$-circuits on $G_{12,3}$.
\end{example}

\section{Orthogonal Fixed-Weight de Bruijn Sequences}
\label{sec:ortho_fixed_w_deB_seq}
As pointed out, in many applications we are allowed to only use sequences with constrained compositions. We therefore first generalize
    the definition of de Bruijn sequences in a way that
    its length-$k$ substrings belong to a constrained subset of length-$k$ sequences.
\begin{definition}\label{def:r_deB_seq}
    \cite[Section 1]{dB_Fix_W_Ruskey}
    Let $k\geq 1$,
        $\mathcal{A}$ be an alphabet of size $\sigma\geq 2$,
        and $L$ be a subset of $\mathcal{A}^k$.
    A circular sequence $\mathbf{s}=(s_0,s_1,\ldots,s_{|L|-1})$ is called
        a \emph{de Bruijn sequence with respect to $L$}
        if each
        sequence in
        $L$
        appears as a
        circular
        substring of $\mathbf{s}$ exactly once.
\end{definition}
Similarly to the nonrestricted case, restricted de Bruijn sequence
    can be characterized as Eulerian circuits of a specialized directed graph,
    defined as follows.
\begin{definition}\label{def:r_deB_graph}
    \cite[Section 2]{dB_Fix_W_Ruskey}
    \cite[Section 3]{DNA_profiles_Kiah}
    Let $k\geq 1$,
        and let $\mathcal{A}$ be an alphabet of size $\sigma\geq 2$. Let $L$ be a proper subset of $\mathcal{A}^k$.
    Define the \emph{de Bruijn graph with respect to $L$} as a directed graph
        $D(L)$
        with the vertex set $V(D(L))$ comprising the collection of all length-$(k-1)$ prefixes
        and length-$(k-1)$ suffixes of sequences in $L$, 
        and with 
        arc set
        \begin{align}
            A(D(L)) \coloneqq
            &\{
                (\mathbf{s}_1,\mathbf{s}_2)\in V(D(L))^2
            ~|~
                \text{there is a word in }L 
                \nonumber\\
                &
                \text{ with prefix }\mathbf{s}_1\text{ and suffix }\mathbf{s}_2
            \}.
        \end{align}
\end{definition}
Then,
    there is a 1-1 correspondence
    between
    de Bruijn sequences with respect to $L$
    and
    Eulerian circuits of $D(L)$. Note that
        for some choices of $L$
        there may not exist a de Bruijn sequence with respect to $L$ or 
        an Eulerian circuit in $D(L)$.

We say that 
    two de Bruijn sequences with respect to $L$ are  \emph{orthogonal}
    if they have no common length-$(k+1)$
    circular substring.
Similarly,
    two de Bruijn sequences with respect to $L$
    are orthogonal iff their corresponding Eulerian circuits on $D(L)$
    are compatible. We seek to analyze the restricted de Bruijn graph $D(L)$
    in Definition~\ref{def:r_deB_seq}
    where $L$ is the collection of all sequences of certain restricted weights.

Formally,
    write $\mathcal{A}=\mathcal{W}\cup\mathcal{X}$,
    where
    $\mathcal{W}$ represents the set of ``weighted symbols,'' while
    $\mathcal{X}$ represents the set of ``nonweighted'' symbols, with both 
    $\mathcal{W}$ and $\mathcal{X}$ nonempty and disjoint.
Then,
    for any $n\geq 1$ and each $\mathbf{s}=(s_0,\ldots,s_{n-1}) \in \mathcal{A}^{n}$,
    we define the weight of $\mathbf{s}$,
    $w(\mathbf{s})$,
    as the number of entries of $\mathbf{s}$ in $\mathcal{W}$:  $w(s) \coloneqq \sum_{i=0}^{n-1} \boldone_{\{{s_i\in\mathcal{W}\}}}.$

Then, for $0\leq w' \leq w \leq k,$ we define 
    $\mathcal{A}^{w}_{w'}(k)$
    to be the collection of all sequences in $\mathcal{A}^k$
    with weight between $w'$ and $w$.
Our interest lies in  
    orthogonal de Bruijn sequences with respect to
    $\mathcal{A}_{w-1}^w(k)$
    for $w\in [1,k]$,
    i.e.,
    compatible Eulerian circuits
    on $D(\mathcal{A}_{w'}^w(k))$
    when
    $w$ and $w'$ differ by $1$.
Note that
    this is the smallest difference 
    between $w$ and $w'$ we can consider
    due to the following fact:
If $w'=w$,
    then
    $D(\mathcal{A}_{w'}^w(k))=D(\mathcal{A}_w^w(k))$ is not strongly connected
    and thus not Eulerian
    unless
    $w\in\{0,1,k-1,k\}$~\cite{dB_Fix_W_Ruskey}.
The reason is that
    any vertex whose weight representation contains the substring $1010$ can never reach a vertex with weight representation $111\cdots 000$.
Here the weight representation of $\mathbf{s}=(s_0,\ldots,s_{k-2})$, denoted by $\bm{\chi}(\mathbf{s})$,
    is the binary vector $\bm{\chi}(\mathbf{s})\coloneqq (\boldone_{s_0\in\mathcal{W}},\ldots,\boldone_{s_{k-2}\in\mathcal{W}}).$

By generalizing the arguments in~\cite{dB_Fix_W_Ruskey},
    it is not hard to see that
    the vertex set $V(D(\mathcal{A}_{w-1}^w(k)))$
    is equal to the set $\mathcal{A}_{w-2}^w(k-1)$.
Then,
    for $\mathbf{s}=(s_0,\ldots,s_{k-2})\in V(D(\mathcal{A}_{w-1}^w(k)))$
    with $w(\mathbf{s})=w-2$,
    we observe that
    its predecessors and successors are
    $\{(\alpha,s_0,\ldots,s_{k-3}) | \alpha\in\mathcal{W}\}$
    and
    $\{(s_1,\ldots,s_{k-2},\beta) | \beta\in\mathcal{W}\}$, respectively.
Thus, the in-degree and out-degree of each $\mathbf{s}\in V(D(\mathcal{A}_{w-1}^w(k)))$
    with $w(\mathbf{s})=w-2$ equals $|\mathcal{W}|$.
Similarly,
    for $\mathbf{s}\in V(D(\mathcal{A}_{w-1}^w(k)))$
    with $w(\mathbf{s})=w$, the in- and out-degrees equal $|\mathcal{X}|$.
For $\mathbf{s}\in V(D(\mathcal{A}_{w-1}^w(k)))$
    with $w(\mathbf{s})=w-1$,
    there are no restrictions,
    and thus
    its in- and out-degrees equal  $|\mathcal{A}|=|\mathcal{W}|+|\mathcal{X}|$.

The following theorem
    characterizes the number of orthogonal de Bruijn sequences with respect to $\mathcal{A}_{w-1}^w(k)$:
\begin{theorem}\label{thm:restr_ortho_deB}
The de Bruijn graph $D(\mathcal{A}_{w-1}^w(k))$
    has a collection of
    $\min(|\mathcal{W}|,|\mathcal{X}|)$
    compatible Eulerian circuits.
Furthermore, this is the largest possible collection. In particular, 
    $\mathcal{A}_{w-1}^w(k)$
    has an Eulerian circuit.
\end{theorem}
This finding is consistent with the results of~\cite[Proposition 3]{DNA_profiles_Kiah},
    which state that
    $D(\mathcal{A}^{w}_{w'}(k))$
    always
    has an Eulerian circuit 
    as long as $w'<w$.
\begin{IEEEproof}[Proof of Theorem~\ref{thm:restr_ortho_deB}]
Without loss of generality, 
   we assume that $|\mathcal{W}|\leq |\mathcal{X}|$
   since the other case can be handled similarly.
Since each vertex in the de Bruijn graph
   $D(\mathcal{A}_{w-1}^w(k))$
   has in-out-degree
   $|\mathcal{W}|$,
   $|\mathcal{X}|$,
   or
   $|\mathcal{W}|+|\mathcal{X}|$,
   there is no collection of
   more than $|\mathcal{W}|$
   pairwise compatible Eulerian circuits
   of $D(\mathcal{A}_{w-1}^w(k))$. 

Before we proceed, we recall the vertex-splitting technique mentioned in~\cite{Compatible_Euler_Fleischner}:
Given a directed graph $G=(V,A)$,
   a vertex $v$ with both in-degree and out-degree $d$,
   and a wiring (transition system) $W$ on $v$,
   one can define another directed graph $G'=(V',A')$
   by ``splitting $v$ along $W$'' as follows:
The vertex set of $G'$ is
   $V'=V\cup\{v_1,v_2,\ldots,v_d\}\setminus\{v\}$.
Each $v_j$ has in-out-degree $1$,
   and its predecessor and successor
   are determined by exactly one edge of $W$.
The arcs on all the other vertices $V\setminus\{v\}$
   remain the same as in $G$.

Then, we enumerate
   $\mathcal{W}=\{w_0,w_1,\ldots,w_{|\mathcal{W}|-1}\}$
   and
   $\mathcal{X}=\{x_0,x_1,\ldots,x_{|\mathcal{X}|-1}\}$.
For each vertex $\mathbf{s}=(s_0,\ldots,s_{k-2})$ of $D(\mathcal{A}_{w-1}^w)$
   of weight $w-2$,
   we can define $|\mathcal{W}|$ edge-disjoint wirings
    $W_0(\mathbf{s}),\ldots,W_{|\mathcal{W}|-1}(\mathbf{s})$
    on $\mathbf{s}$ by the following rule:
For $j\in [0,|\mathcal{W}|-1]$,
   the wiring $W_j(\mathbf{s})$ is defined by
   pairing the in-arc
   $(w_i,s_0,\ldots,s_{k-3})$
   with
   the out-arc 
   $(s_1,\ldots,s_{k-2},w_{(i+j)\mod |\mathcal{W}|})$
    for each $i\in [0,|\mathcal{W}|-1]$.
Similarly,
   for each vertex $\mathbf{v}=(v_0,\ldots,v_{k-2})$ in  $D(\mathcal{A}_{w-1}^w)$ of  weight $w$
   and for $j\in [0,|\mathcal{W}|-1]$,
   we define the wiring $W'_j(\mathbf{v})$ by
   pairing the in-edge
   $(x_i,s_0,\ldots,s_{k-3})$
   with
   the out-edge
   $(s_1,\ldots,s_{k-2},x_{(i+j)\mod |\mathcal{X}|})$
   for each $i\in[0,|\mathcal{X}|-1]$.
Recall that $|\mathcal{W}|\leq|\mathcal{X}|$,
   and thus $W'_0(\mathbf{v}),\ldots,W'_{|\mathcal{W}|-1}(\mathbf{v})$
   are edge-disjoint as well.
 
Now,
   for each $j\in [0,|\mathcal{W}|-1]$,
   we define
   $G_j$
   to be the directed graph obtained from $D(\mathcal{A}_{w-1}^w(k))$ by
   splitting \emph{every} vertex $\mathbf{s}$ of weight $w-2$ along $W_j(\mathbf{s})$
   and
   splitting \emph{every} vertex $\mathbf{v}$ of weight $w$ along $W'_j(\mathbf{v})$.
If each $G_j$ admits an Eulerian cycle,
   then by ``merging back the splitted vertices'',
   we obtain $|\mathcal{W}|$
   Eulerian circuits $C_0,\ldots,C_{|\mathcal{W}|-1}$ on $D(\mathcal{A}_{w-1}^w(k))$
   such that
   for each vertex with weight $w-2$ or $w$
   no in-out arc-pair is used twice in these $C_j$.
Then,
   since the remaining weight-$(w-1)$ vertices
   have degree
   $|\mathcal{W}|+|\mathcal{X}|\geq 2|\mathcal{W}|$,
   for each $j$
   we can repeatedly apply Lemma~\ref{lem:rewiring_general}
   to rewire each weight-$(w-1)$ vertex in $C_j$
   given $C_{0},\ldots,C_{j-1}$.
After this process,
   we obtain $|\mathcal{W}|$
   compatible Eulerian circuits
   on $D(\mathcal{A}_{w-1}^w(k))$.

It remains to be shown that each $G_j$ indeed has an Eulerian circuit.
Since each vertex in $G_j$ either has in-out degree $1$
   or in-out degree $|\mathcal{W}|+|\mathcal{X}|$,
   it suffices to verify that $G_j$ is strongly connected.
Let $\mathbf{v}=(v_0,\ldots,v_{k-2})$ and $\mathbf{v}'=(v'_0,\ldots,v'_{k-2})$ be two  vertices in $G_j$.
We first show that from $\mathbf{v}$ we can reach $\mathbf{v}'$ if $\mathbf{v}'$ has weight $w-1$.
Write the weight representations of $\mathbf{v}$ and $\mathbf{v}'$ as
   $\bm{\chi}(\mathbf{v})=(b_0,\ldots,b_{k-2})$
   and
   $\bm{\chi}(\mathbf{v}')=(b'_0,\ldots,b'_{k-2})$,
   respectively.
Also,
   consider
   the following generalized de Bruijn graph
   $D(\mathcal{B}_{w-1}^w(k))$,
   where $\mathcal{B}$ consists of
   only one weighted symbol $1$
   and
   one nonweighted symbol $0$.
By~\cite[Corollary 2.3]{dB_Fix_W_Ruskey},
   there is a path $P$ from $\bm{\chi}(\mathbf{v})$ to $\bm{\chi}(\mathbf{v}')$
   on
   $D(\mathcal{B}_{w-1}^w(k))$.
Write the binary sequence induced by $P$ as
   $\mathbf{p}=(a_0,\ldots a_T)$,
   where $(a_0,\ldots, a_{k-2})=(b_0,\ldots, b_{k-2})$
   and
   $(a_{T-k+2},\ldots, a_T)=(b'_{0},\ldots, b'_{k-2}$.
Then,
   observe that
   we can ``follow the same path $P$''
   to walk on $G_j$ from $\mathbf{v}$ to some vertex having the same weight representation 
   as $\mathbf{v}'$.
That is,
   there is a path $P'$ on $G_j$ from $\mathbf{v}$ to some $\mathbf{v}''=(v''_0,\ldots,v''_{k-2})$
   such that
   the induced sequence of $P'$ has weight representation $\mathbf{p}$
   and
   the weight representation of $\mathbf{v}''$
   satisfies
   $\bm{\chi}(\mathbf{v}'')=\bm{\chi}(\mathbf{v}')$.
This claim can be easily established by induction on the length of $P$.

Then,
   note that
   since $\mathbf{v}''$ and $\mathbf{v}'$ have the same weight representation
   and they all share the same weight $w-1$,
   we can traverse from $\mathbf{v}''$ to $\mathbf{v}'$
   using a walk $P''$ that induces the sequence
   $\mathbf{p}''=(v''_0,\ldots,v''_{k-2},v'_0,\ldots,v'_{k-2})$. 
The weight representation
   of each length-$k$ substring in $\mathbf{p}''$
   has a prefix that is 
   a circular shift of $\bm{\chi}(\mathbf{v}')$,
   and thus
   each length-$k$ sequence in $\mathbf{p}''$
   has weight $w$ or $w-1$.
Therefore,
   $P''$ is a valid walk in $D(\mathcal{A}_{w-1}^w(k))$.
Furthermore,
   each length-$(k-1)$ substring in $\mathbf{p}''$
   has a weight representation equal to a circular shift of $\bm{\chi}(\mathbf{v}')$,
   and thus the walk $P''$ in $D(\mathcal{A}_{w-1}^w(k))$
   only
   visits
   vertices of weight $w-1$.
Thus,
   the walk $P''$ is unaffected by the splitting process
   and remains present in $G_j$.
In conclusion,
   one can walk on $G_j$ from $\mathbf{v}$ to $\mathbf{v}''$ and then
   from $\mathbf{v}''$ to $\mathbf{v}'$,
   provided that $\mathbf{v}'$ has weight $w-1$.

Now consider the case where
    $\mathbf{v}'$ has weight $w$ or $w-2$.
If $\mathbf{v}'$ has weight $w-2$,
   we can traverse in a reverse direction
   from $\mathbf{v}'$ to its unique predecessor $\text{pred}(\mathbf{v}')$,
   and then if $\text{pred}(\mathbf{v}')$ still has weight $w-2$,
   traverse to $\text{pred}(\text{pred}(\mathbf{v}'))$,
   and so on.
Since each traversal appends a symbol of weight $1$
   to the left of $\mathbf{v}'$ and removes the right-most symbol of $\mathbf{v}'$,
   we will reach a weight-$(w-1)$ vertex (say $\hat{\mathbf{v}}$) in $G_j$
   after $k-\tau+1$ reverse steps,
   where $b'_{\tau}$ is the last $0$ in $\bm{\chi}(\mathbf{v}')$.
We can then apply the previous result to see that we can start  from $\mathbf{v}$ to reach $\hat{\mathbf{v}}$,
   and then, since $\hat{\mathbf{v}}$ is obtained from 
   a reverse traversal from $\mathbf{v}'$,
   walk from $\hat{\mathbf{v}}$ to $\mathbf{v}'$.
A similar argument holds
    when
    $\mathbf{v}'$ has weight $w$.

These arguments show that $G_j$ is strongly connected
   and thus has an Eulerian circuit.
\end{IEEEproof}
\begin{example}
Consider
   $\mathcal{W}=\{C,G\}$,
   $\mathcal{X}=\{A,T\}$,
   $k=4$,
   and $w=3$.
Then, 
   $\mathcal{A}=\{A,T,C,G\}$
   and
   $\mathcal{A}_{w-1}^w(k)=\mathcal{A}_{2}^3(4)$
 represent the collection of all length-$4$
   sequences of $\{A,T,C,G\}$
   with weight $2$ or $3$,
   where the ``weight'' of a word
   is determined by the number of symbols that are either $C$ or $G$ (i.e., the $GC$ content).
The vertices of 
   the de Bruijn graph
   $D(\mathcal{A}_{2}^3(4))$
   are all length-$3$ sequences of weight $1$, $2$, or $3$.
For example,
   consider the vertex
   $CAA$,
   which has weight $1$.
It has two in-arcs,
   one from $CCA$
   and the other from $GCA$.
The two out-arcs of this vertex
   point towards 
   $AAC$ and $AAG$,
   respectively.
   
For each vertex $\tilde{\mathbf{v}}$ of
   $D(\mathcal{A}_{2}^3(4))$
   with weight $1$ or $3$,
   the wiring $W_0(\tilde{\mathbf{v}})$
   associates each of its predecessor $\mathbf{p}$
   to the successor $\mathbf{s}$
   such that the first symbol of $\mathbf{p}$ equals the last symbol of $\mathbf{s}$.
For example,
   the vertex $\hat{\mathbf{v}}=CAA$
   has predecessors
   $\mathbf{p}_0=CCA,\mathbf{p}_1=GCA$
   and successors
   $\mathbf{s}_0=AAC,\mathbf{s}_1=AAG$,
   and
   the wiring $W_0(\hat{\mathbf{v}})$
   pairs $\mathbf{p}_0$ with $\mathbf{s}_0$
   and $\mathbf{p}_1$ with $\mathbf{s}_1$.
Then,
   in the split graph $G_0$,
   the original vertex $\hat{\mathbf{v}}=CAA$
   is split to 
   two vertices $CAA_C$ and $CAA_G$,
   where the subscript $C$ or $G$
   denotes the first symbol of its only predecessor,
   which is equal to the last symbol of its only successor.
More explicitly,
   $CAA_C$ has only one in-arc from $CCA$ and only one out-arc  towards $AAC_C$,
   and
   $CAA_G$ has only one in-arc from $GCA$ and only one out-arc  towards $AAG_C$.

We show next that from $\mathbf{v}=CAA_G$ we can for example reach 
   $\mathbf{v}^*=TCT_G$
   on this split graph $G_0$.
First,
   since $\mathbf{v}^*=TCT_G$ has weight $1=w-2$,
   we first reversely traverse the graph starting from $TCT_G$ to 
   reach its only predecessor
   $\mathbf{v}'=GTC$,
   which has weight $2=w-1$.
Then,
   the weight representations
   of $\mathbf{v}$ and $\mathbf{v}'$ read as 
   $100$ and $101$,
   respectively.
We then
    apply the proof steps in~\cite[Lemma 2.2]{dB_Fix_W_Ruskey}
    and~\cite[Corollary 2.3]{dB_Fix_W_Ruskey}
    to construct
   a path $P$ from $100$ to $101$
   on the binary-version de Bruijn graph $D(\mathcal{B}_2^3(4))$.
It turns out that
    this path $P$
    induces the sequence $1001101$
    on $D(\mathcal{B}_2^3(4))$,
    and
    by an abuse of notation
    we write $P=1001101$.
Based on $P$,
   we find that
   the path
   $P'=CAAGCAC$
   is a valid path on $G_0$
   from $\mathbf{v}=CAA_G$ to $\mathbf{v}''=CAC$.
More explicitly,
   $P'$ traverses in the following order: $CAA_G\rightarrow AAG_C\rightarrow AGC \rightarrow GCA \rightarrow CAC$.
Note that the weight representation of $\mathbf{v}''=CAC$
   is the same as that of 
   $\mathbf{v}'=GTC$.
Then,
   we can initiate a walk from $\mathbf{v}''=CAC$
   to $\mathbf{v}'=GTC$
   following the path
   $CACGTC$
   on $G_0$.
These arguments shows that from 
   $\mathbf{v}$ we can reach $\mathbf{v}^*$ through 
   $\mathbf{v}\rightarrow \mathbf{v}''\rightarrow \mathbf{v}' \rightarrow \mathbf{v}^*$.
\end{example}

\section{Generalized Orthogonal Kautz Sequences}
Next, we study possible extensions of the concepts of  generalized orthogonal de Bruijn sequences to Kautz sequences. More explicitly, we introduce the notions of $\ell$-orthogonal Kautz sequences, orthogonal balanced Kautz sequences, and orthogonal fixed-weight Kautz sequences. We start by recalling the definition of Kautz sequences.

\begin{definition}\label{def:Kautz_seq}
    Let $k\geq 1$ and let $\mathcal{A}$ be an alphabet of size $\sigma\geq 3$. Define $\mathcal{K}_k(\mathcal{A})$ to be the collection of all length-$k$ strings in $\mathcal{A}^k$ that do 
        not have two or more adjacent identical characters (i.e., no runlengths/homopolymers of length longer than two).
    Then,
        a circular sequence $\mathbf{s}=(s_0,s_1,\ldots,s_{\sigma(\sigma-1)^{k-1}-1})$ is called
        a \emph{$(\sigma,k)$-Kautz sequence}
        if each 
        sequence in
        $\mathcal{K}_k(\mathcal{A})$
        appears as a
        circular
        substring of $\mathbf{s}$ exactly once.
\end{definition}
\begin{remark}
    Using the terminology in Definition \ref{def:r_deB_seq},
        a Kautz sequence can be interpreted as
        a de Bruijn sequence with respect to
        the set $\mathcal{K}_k(\mathcal{A})$.
\end{remark}

Similarly as de Brujin sequences, Kautz sequences are closely related to \emph{Kautz graphs}.
Using the notation in Definition \ref{def:r_deB_graph},
a Kautz graph 
    $G^{\textnormal{Kautz}}_{\sigma,k}\coloneqq D(\mathcal{K}_k(\mathcal{A}))$ is a generalized de Bruijn graph with respect to the set $\mathcal{K}_k(\mathcal{A})$.
Consequently,
    there is a one-to-one correspondence 
    between
    the $(\sigma,k)$-Kautz sequences 
    and 
    the Eulerian circuits on $G^{\textnormal{Kautz}}_{\sigma,k}$
    \cite{Ortho_dB_Skiena}.
In addition,
    there is also a one-to-one correspondence 
    between
    the $(\sigma,k)$-Kautz sequences 
    and 
    the Hamiltonian cycles on $G^{\textnormal{Kautz}}_{\sigma,k+1}$
    \cite{Ortho_dB_Skiena}.

\begin{example}
Consider the DNA alphabet $\mathcal{A}=\{A,T,C,G\}$
    and $k=2$.
The sequence $\mathbf{s}=ATCGAGCTGTAC$ is a $(4,2)$-Kautz sequence
    since
    each length-$2$ sequence of unequal symbols from $\mathcal{A}$
    appears as a circular substring of $\mathbf{s}$ exactly once.
Fig.~\ref{fig:Kautz_4_2_4_3}\subref{subfig:Kautz_4_2}
    depicts the Kautz graph $G^{\textnormal{Kautz}}_{4,2}$,
    which has an Eulerian circuit corresponding to $\mathbf{s}$.
The wiring induced by that circuit
    is illustrated in 
    Fig.~\ref{fig:Kautz_4_2_4_3}\subref{subfig:Kautz_4_2_wired}.
The sequence $\mathbf{s}$
    can also be represented by
    a Hamiltonian cycle on the Kautz graph
    $G^{\textnormal{Kautz}}_{4,3}$
    in Fig.~\ref{fig:Kautz_4_2_4_3}\subref{subfig:Kautz_4_3}.
This cycle is given by
                $AT \rightarrow TC \rightarrow CG 
     \rightarrow GA \rightarrow AG \rightarrow GC
     \rightarrow CT \rightarrow TG \rightarrow GT
     \rightarrow TA \rightarrow AC \rightarrow CA
     \rightarrow AT
    $.
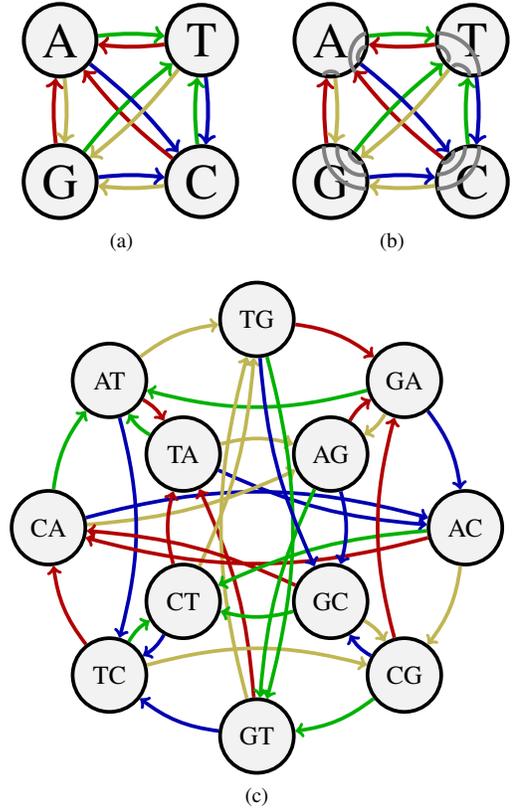
\begin{figure}
    \centering
    \subfloat[][]{
        \scalebox{0.94}{
            \ifdefined\DrawTikz
                \newcommand{\LD}{\langle}
\newcommand{\RD}{\rangle}
\begin{tikzpicture}[scale=1]

\tikzset{vertex/.style = {draw, circle, color=black, fill=black!5, ultra thick,minimum size=1cm}}
\tikzset{arcA/.style = {->, ultra thick, color = red!70!black}}
\tikzset{arcT/.style = {->, ultra thick, color = green!70!black}}
\tikzset{arcC/.style = {->, ultra thick, color = blue!70!black}}
\tikzset{arcG/.style = {->, ultra thick, color = yellow!70!black}}
\tikzset{wiring/.style = {gray, ultra thick}}

\node[vertex] (A) at (-1,1) {\huge A};
\node[vertex] (T) at (1,1) {\huge T};
\node[vertex] (C) at (1,-1) {\huge C};
\node[vertex] (G) at (-1,-1) {\huge G};

\draw[arcA] (T) to [bend left=6.5] (A);
\draw[arcA] (C) to [bend left=6.5] (A);
\draw[arcA] (G) to [bend left=6.5] (A);
\draw[arcT] (A) to [bend left=6.5] (T);
\draw[arcT] (C) to [bend left=6.5] (T);
\draw[arcT] (G) to [bend left=6.5] (T);
\draw[arcC] (A) to [bend left=6.5] (C);
\draw[arcC] (T) to [bend left=6.5] (C);
\draw[arcC] (G) to [bend left=6.5] (C);
\draw[arcG] (A) to [bend left=6.5] (G);
\draw[arcG] (T) to [bend left=6.5] (G);
\draw[arcG] (C) to [bend left=6.5] (G);


\end{tikzpicture}%
            \fi
        }
        \label{subfig:Kautz_4_2}
    }
    \subfloat[][]{
        \scalebox{0.94}{
            \ifdefined\DrawTikz
                \newcommand{\LD}{\langle}
\newcommand{\RD}{\rangle}
\begin{tikzpicture}[scale=1]

\tikzset{vertex/.style = {draw, circle, color=black, fill=black!5, ultra thick,minimum size=1cm}}
\tikzset{arcA/.style = {->, ultra thick, color = red!70!black}}
\tikzset{arcT/.style = {->, ultra thick, color = green!70!black}}
\tikzset{arcC/.style = {->, ultra thick, color = blue!70!black}}
\tikzset{arcG/.style = {->, ultra thick, color = yellow!70!black}}
\tikzset{wiring/.style = {gray, ultra thick}}

\node[vertex] (A) at (-1,1) {\huge A};
\node[vertex] (T) at (1,1) {\huge T};
\node[vertex] (C) at (1,-1) {\huge C};
\node[vertex] (G) at (-1,-1) {\huge G};

\draw[arcA] (T) to [bend left=6.5] (A);
\draw[arcA] (C) to [bend left=6.5] (A);
\draw[arcA] (G) to [bend left=6.5] (A);
\draw[arcT] (A) to [bend left=6.5] (T);
\draw[arcT] (C) to [bend left=6.5] (T);
\draw[arcT] (G) to [bend left=6.5] (T);
\draw[arcC] (A) to [bend left=6.5] (C);
\draw[arcC] (T) to [bend left=6.5] (C);
\draw[arcC] (G) to [bend left=6.5] (C);
\draw[arcG] (A) to [bend left=6.5] (G);
\draw[arcG] (T) to [bend left=6.5] (G);
\draw[arcG] (C) to [bend left=6.5] (G);

\draw[wiring] (0.5,1.1) to [bend left=45] (1.1,0.5); 
\draw[wiring] (1.1,-0.5) to [bend left=45] (0.5,-1.1); 
\draw[wiring] (-0.5,-1.1) to [bend left=45] (-1.1,-0.5); 
\draw[wiring] (-1.1,0.5) to [bend left=90] (-0.9,0.5); 
\draw[wiring] (-0.9,-0.5) to [bend right=60] (-0.5,-0.9); 
\draw[wiring] (0.5,-0.9) to [bend right=45] (0.9,-0.5); 
\draw[wiring] (0.95,0.5) to [bend right=90] (0.725,0.625); 
\draw[wiring] (-0.55,-0.72) to [bend left=90] (-0.72,-0.55); 
\draw[wiring] (0.65,0.7) to [bend right=90] (0.5,0.9); 
\draw[wiring] (-0.5,0.9) to [bend right=90] (-0.59,0.7); 
\draw[wiring] (0.72,-0.57) to [bend left=90] (0.57,-0.72); 
\draw[wiring] (-0.65,0.55) to [bend left=60] (-0.5,1.1); 

\end{tikzpicture}%
            \fi
        }
        \label{subfig:Kautz_4_2_wired}
    }\\
    \subfloat[][]{
        \scalebox{0.49}{
            \ifdefined\DrawTikz
                \begin{tikzpicture}[scale=1]

\tikzset{vertex/.style = {draw, circle, color=black, fill=black!5, line width=0.1cm,minimum size=2cm}}
\tikzset{arcA/.style = {->, line width=0.1cm, color = red!70!black}}
\tikzset{arcT/.style = {->, line width=0.1cm, color = green!70!black}}
\tikzset{arcC/.style = {->, line width=0.1cm, color = blue!70!black}}
\tikzset{arcG/.style = {->, line width=0.1cm, color = yellow!70!black}}
\tikzset{BarcA/.style = {->, line width=0.1cm, color = red!70!black}}
\tikzset{BarcT/.style = {->, line width=0.1cm, color = green!70!black}}
\tikzset{BarcC/.style = {->, line width=0.1cm, color = blue!70!black}}
\tikzset{BarcG/.style = {->, line width=0.1cm, color = yellow!70!black}}

\node[vertex] (AT) at (-4,4) {\huge AT};
\node[vertex] (AC) at (5.656,0) {\huge AC};
\node[vertex] (AG) at (2,2) {\huge AG};
\node[vertex] (TA) at (-2,2) {\huge TA};
\node[vertex] (TC) at (-4,-4) {\huge TC};
\node[vertex] (TG) at (0,5.656) {\huge TG};
\node[vertex] (CA) at (-5.656,0) {\huge CA};
\node[vertex] (CT) at (-2,-2) {\huge CT};
\node[vertex] (CG) at (4,-4) {\huge CG};
\node[vertex] (GA) at (4,4) {\huge GA};
\node[vertex] (GT) at (0,-5.656) {\huge GT};
\node[vertex] (GC) at (2,-2) {\huge GC};

\draw[arcT] (TA) to [bend left=15] (AT);
\draw[BarcT] (CA) to [bend left=15] (AT);
\draw[arcT] (GA) to [bend left=15] (AT);
\draw[arcC] (CA) to [bend left=15] (AC);
\draw[BarcC] (TA) to [bend right=10] (AC);
\draw[arcC] (GA) to [bend left=15] (AC);
\draw[arcG] (TA) to [bend left=15] (AG);
\draw[arcG] (CA) to [bend right=10] (AG);
\draw[BarcG] (GA) to [bend left=15] (AG);

\draw[arcA] (AT) to [bend left=15] (TA);
\draw[arcA] (CT) to [bend left=15] (TA);
\draw[BarcA] (GT) to [bend right=10] (TA);
\draw[BarcC] (AT) to [bend left=15] (TC);
\draw[arcC] (CT) to [bend left=15] (TC);
\draw[arcC] (GT) to [bend left=15] (TC);
\draw[arcG] (AT) to [bend left=15] (TG);
\draw[BarcG] (CT) to [bend right=10] (TG);
\draw[arcG] (GT) to [bend left=15] (TG);

\draw[BarcA] (AC) to [bend left=15] (CA);
\draw[arcA] (TC) to [bend left=15] (CA);
\draw[arcA] (GC) to [bend right=10] (CA);
\draw[arcT] (AC) to [bend right=10] (CT);
\draw[arcT] (TC) to [bend left=15] (CT);
\draw[BarcT] (GC) to [bend left=15] (CT);
\draw[arcG] (AC) to [bend left=15] (CG);
\draw[BarcG] (TC) to [bend left=15] (CG);
\draw[arcG] (GC) to [bend left=15] (CG);

\draw[arcA] (AG) to [bend left=15] (GA);
\draw[BarcA] (CG) to [bend left=15] (GA);
\draw[arcA] (TG) to [bend left=15] (GA);
\draw[arcT] (AG) to [bend right=10] (GT);
\draw[arcT] (CG) to [bend left=15] (GT);
\draw[BarcT] (TG) to [bend left=15] (GT);
\draw[BarcC] (AG) to [bend left=15] (GC);
\draw[arcC] (CG) to [bend left=15] (GC);
\draw[arcC] (TG) to [bend right=10] (GC);


\end{tikzpicture}%
            \fi
        }
        \label{subfig:Kautz_4_3}
    }
    \caption{
        \protect\subref{subfig:Kautz_4_2}
            The Kautz graph $G^{\textnormal{Kautz}}_{4,2}$.
        \protect\subref{subfig:Kautz_4_2_wired}
            The wiring of $G^{\textnormal{Kautz}}_{4,2}$
                induced by
                the Eulerian circuit
                corresponding to
                the $(4,2)$-Kautz sequence
                ATCGAGCTGTAC.
        \protect\subref{subfig:Kautz_4_3}
            The Kautz graph $G^{\textnormal{Kautz}}_{4,3}$.
        }
    \label{fig:Kautz_4_2_4_3}
\end{figure}
\end{example}

\subsection{$\ell$-Orthogonal Kautz Sequences}
By generalizing Definition~\ref{def:ell_ortho},
    we describe the notion of $\ell$-orthogonality of Kautz sequences
    as follows. 
A collection of $(\sigma,k)$-Kautz sequences $\mathcal{C}=\{\mathbf{s}_1,\ldots,\mathbf{s}_M\}$
    is called $\ell$-orthogonal
    if each length-$(k+1)$ string in $\mathcal{K}_{k+1}(\mathcal{A})$
    appears at most $\ell$ times in $\mathcal{C}$,
    where $\ell \geq 1$
    and $\mathcal{A}$ is an alphabet of size $\sigma$.
Similar to the case of de Bruijn sequences,
    for any collection $\mathcal{C}$ of $(\sigma,k)$-Kautz sequences,
    the following statements are equivalent:
\begin{itemize}
    \item 
        $\mathcal{C}$ is $\ell$-orthogonal.
    \item 
        The corresponding Eulerian circuits on $G^{\textnormal{Kautz}}_{\sigma,k}$
            use
            each pair of consecutive arcs
            at most $\ell$ times.
    \item 
        The corresponding Hamiltonian cycles on $G^{\textnormal{Kautz}}_{\sigma,k+1}$
            use
            each arc
            at most $\ell$ times.
\end{itemize}

Let $\Omega^{\textnormal{Kautz}}_{\ell}(\sigma,k)$ be the maximum cardinality of
    any collection of ${\ell}$-orthogonal
    $(\sigma,k)$-Kautz sequences.
By adapting the proofs of Proposition~\ref{prop:ell_ortho_ub}
    and Theorem~\ref{thm:ell_ortho_lb},
    we can bound  $\Omega^{\textnormal{Kautz}}_{\ell}(\sigma,k)$ as follows.

\begin{proposition}\label{prop:ell_ortho_Kautz_bds}
    We have $\Omega^{\textnormal{Kautz}}_{\ell}(\sigma,k)\leq \ell(\sigma-1)$.
    If $\ell \leq \sigma(\sigma-1)^{k-2}$
        and $\sigma \geq 4$,
        then we further have
        $\Omega^{\textnormal{Kautz}}_{\ell}(\sigma,k)\geq 2\ell$
        and
        $\Omega^{\textnormal{Kautz}}_{\ell}(\sigma,k)\geq \ell\lfloor (\sigma-1)/2\rfloor$.
\end{proposition}
\begin{IEEEproof}[Skecth of the proof]
First observe that
    every vertex in the Kautz graph $G^{\textnormal{Kautz}}_{\sigma,k}$
    has
    $\sigma-1$ inward arcs,
    $\sigma-1$ outward arcs,
    and no loop.
The same arguments in the proof of Proposition \ref{prop:ell_ortho_ub}
    proves the upper bound on  $\Omega^{\textnormal{Kautz}}_{\ell}(\sigma,k)$.

Furthermore,
    note that
    by adapting the proof of Theorem \ref{thm:ell_ortho_lb},
    we actually can establish the following claim: 
For any Eulerian directed graph $G$
    with minimum in-out-degree $\delta$
    and any
    $\ell\leq |V(G)|$,
    $G$
    has a collection of $\ell K$ Eulerian circuits
    such that
    each pair of consecutive arcs
    appears in this collection at most $\ell$ times,
    where $K=2$ for $\delta=3$ and $K=\lfloor \delta /2 \rfloor$ for $\delta \geq 4$.
The lower bounds on  $\Omega^{\textnormal{Kautz}}_{\ell}(\sigma,k)$ can then be  immediately deduced by the fact that the minimum in-out-degree of $G^{\textnormal{Kautz}}_{\sigma,k}$ is $\sigma-1$.
\end{IEEEproof}

\begin{example}
We verify
    that
    the lower bound
    of $\Omega_{\ell}^{\textnormal{Kautz}}(\sigma,k)$
    in Proposition \ref{prop:ell_ortho_Kautz_bds}
    holds
    for $\sigma=4$,
    $k=2$,
    and $\ell=2$.
Setting these parameters in
    Proposition \ref{prop:ell_ortho_Kautz_bds}
    gives
    $\Omega_{\ell}^{\textnormal{Kautz}}(\sigma,k)\geq 2\ell=4$.
Thus,
    we seek to find
    a collection of
    four
    $(4,2)$-Kautz sequences
    such that
    each length-$3$ string in $\mathcal{K}_{3}(\mathcal{A})$
    appears at most twice in $\mathcal{C}$.
Following the proof of Theorem \ref{thm:ell_ortho_lb},
    we
    partition
    $V(G_{4,2})$ into
    $\mathcal{V}_1=\{A,T\}$
    and
    $\mathcal{V}_2=\{C,G\}$
    and
    start with the Eulerian circuit $C_{1,1}=ATCGAGCTGTAC$
    on $G_{4,2}$.
After the rewiring process,
    we obtain the following three more $(4,2)$-Kautz sequences:
\begin{align*}
    &C_{1,2}=ACAGCTATGTCG,
    &C_{2,1}=ACTATGCGTCAG,\\
    &C_{2,2}=ACTGCGTAGATC.
\end{align*}
It can be checked that
    the collection $\mathcal{C}=\{C_{1,1},C_{1,2},C_{2,1},C_{2,2}\}$
    is $2$-orthogonal.
For example,
    the string $ATC$ appears twice
    (in $C_{1,1}$ and in $C_{2,2}$), while 
    the string $GAG$ appears once (in $C_{1,1}$),
    and the string $ATA$ does not appear at all.
\end{example}

\subsection{Orthogonal Balanced Kautz Sequences}
We now generalize Definition \ref{def:bal_deB} in order to define balanced Kautz sequences as follows.
\begin{definition}\label{def:bal_Kautz}
    Let $k$ and $\mathcal{A}$ be as before and let $b\geq 1$.
    We say that a circular sequence $\mathbf{s}=(s_0,s_1,\ldots,s_{b\sigma(\sigma-1)^{k-1}-1})$
        is a $b$-balanced \emph{$(\sigma,k)$-Kautz sequence}
        if each sequence in
        $\mathcal{K}_k(\mathcal{A})$
        appears as a
        circular
        substring of $\mathbf{s}$ exactly $b$ times. 
\end{definition}
The following terminology is similar to that used in Section \ref{sec:ortho_bal_deB}.
We say that a collection $\mathcal{C}$ of $b$-balanced $(\sigma,k)$-Kautz sequences is \emph{orthogonal}
    if each sequence in $\mathcal{K}_{k+1}(\mathcal{A})$
    appears at most once in $\mathcal{C}$.
A necessary condition for a $(\sigma,k)$-Kautz sequence $\mathbf{s}$ to belong to an orthogonal collection is that
    $\mathbf{s}$ itself contains no sequence in $\mathcal{K}_{k+1}(\mathcal{A})$ more than once.
In this case we say that $\mathbf{s}$ is self-orthogonal.
Then,
    there is a one-to-one correspondence between
    self-orthogonal $b$-balanced $(\sigma,k)$-Kautz sequences
    and $b$-circuits on $G^{\textnormal{Kautz}}_{\sigma,k+1}$,
    where the $b$-circuits are defined as in Definition~\ref{def:b_ckt}.

For $c\geq 2$,
    we define $\Sigma^{\textnormal{Kautz}}(c,b,k)$
    to be the smallest $\sigma$
    such that
    there exist $c$ orthogonal $b$-balanced $(\sigma,k)$-Kautz sequences.
We have the following bounds of $\Sigma^{\textnormal{Kautz}}(c,b,k)$:
\begin{proposition}\label{prop:ortho_bal_Kautz_bds}
Assume $b\geq 2$ and $c\geq 2$. Then we have
    $cb+1\leq \Sigma^{\textnormal{Kautz}}(c,b,k) \leq 2cb+1$.
\end{proposition}
\begin{IEEEproof}[Skecth of the proof]
The proof of the lower bound of $\Sigma^{\textnormal{Kautz}}(c,b,k)$
    is similar to that of Proposition \ref{prop:Sigma_lb}
    and is omitted.

To prove the upper bound,
    it suffices to prove that $G^{\textnormal{Kautz}}_{2cb+1,k}$ has
    $c$ arc-disjoint $b$-circuits.
We first
    invoke 
    Lemma \ref{lem:rewiring_general}
    to deduce that
    there are $\lfloor (2cb+1-1)/2\rfloor = cb$
    pairwise compatible Eulerian circuits on $G^{\textnormal{Kautz}}_{2cb+1,k}$.
These circuits
    correspond to
    $cb$ arc-disjoint Hamiltonian cycles on $G^{\textnormal{Kautz}}_{2cb+1,k}$.
Then,
    we can arbitrary partition
    the collection of these cycles
    into $c$ groups,
    each having $b$ cycles.
The combination of the cycles in each group
    is a $b$-circuit on $G^{\textnormal{Kautz}}_{2cb+1,k}$,
    and the $b$-circuits combined from different groups are arc-disjoint.
Therefore,
    $G^{\textnormal{Kautz}}_{2cb+1,k}$ has
    $c$ arc-disjoint $b$-circuits,
    which proves the upper bound on  $\Sigma^{\textnormal{Kautz}}(c,b,k)$.
\end{IEEEproof}
\begin{remark}
Since de Bruijn sequences
    are also characterized by both Eulerian circuits and Hamiltonian cycles,
    applying the same steps in the proof of Proposition \ref{prop:ortho_bal_Kautz_bds}
    gives
    $\Sigma(c,b,k)\leq 2cb$,
    where $\Sigma(c,b,k)$ is defined in
    Section \ref{sec:ortho_bal_deB}.
However,
    Theorem \ref{thm:Sigma_ub} always gives a upper bound
    of $\Sigma(c,b,k)$
    strictly smaller than $2cb$
    since
    there is always a power of two contained in $[cb,2cb-1]$.
\end{remark}

\subsection{Orthogonal Fixed-Weight Kautz Sequences}
We first define fixed-weight Kautz sequences.
Following the definitions in Section \ref{sec:ortho_fixed_w_deB_seq},
    consider the case $\mathcal{A}=\mathcal{W}\cup\mathcal{X}$
    with each symbol in $\mathcal{W}$ having weight $1$
    and each symbol in $\mathcal{X}$ having zero weight.
For $0\leq w' \leq w \leq n$,
    we then define a fixed-weight Kautz sequence 
    with parameters
    $(|\mathcal{A}|,k,w',w)$
    as a de Bruijn sequence with respect to $\mathcal{K}_k(\mathcal{A})\cap \mathcal{A}_{w'}^w(k)$. 
We have the following conditions on the existence of such sequences:
\begin{proposition}\label{prop:fixed_w_Kautz_exist}
A fixed-weight Kautz sequence
    with parameters $(|\mathcal{A}|,k,w',w)$
    exists
    if and only if $w'$ and $w$ satisfy one of the following conditions:
\begin{enumerate}
    \item $w'=w=0$,
    \item $w'=w=k$,
    \item $w'\in \{0,1\}$ and $w\in\{k-1,k\}$.
\end{enumerate}
\end{proposition}
\begin{IEEEproof}
We first prove that
    the conditions in Proposition \ref{prop:fixed_w_Kautz_exist}
    imply the existence of a fixed-weight Kautz sequence
    with associated parameters.
If $w'=w=0$,
    then $\mathcal{K}_k(\mathcal{A})\cap\mathcal{A}_{w'}^w(k)=\mathcal{K}_k(\mathcal{X})$,
    and the fixed-weight sequences are simply the $(|\mathcal{X}|,k)$-Kautz sequences.
Similarly,
    the case $w'=w=k$,
    corresponds to $(|\mathcal{W}|,k)$-Kautz sequences.
Among the four subcases with  $w'\in\{0,1\}$ and $w\in\{k-1,k\}$,
    the subcase $(w',w)=(0,k)$
    corresponds to $(|\mathcal{A}|,k)$-Kautz sequences
    since $\mathcal{K}_k(\mathcal{A})\cap\mathcal{A}_{w'}^w(k)=\mathcal{K}_k(\mathcal{A})$.
The remaining three subcases $(w',w)=(0,k-1)$, $(w',w)=(1,k-1)$, and $(w',w)=(1,k)$ can be seen as a direct consequence of Proposition ~\ref{prop:fixed_w_Kautz_ortho},
    which we will state and prove later.

Now assume none of the conditions in Proposition \ref{prop:fixed_w_Kautz_exist} hold.
We seek to show that
    the generalized de Bruijn graph
    $D(\mathcal{K}_k(\mathcal{A})\cap\mathcal{A}_{w'}^w(k))$
    is not Eulerian.
Since Condition 3 fails to hold,
    we have either $w'\geq 2$ or $w\leq k-2$.
First assume $w'\geq 2$.
Since Condition 2 fails to hold 
    and $w'\leq w$,
    we can assume $w'\leq k-1$.
Now,
    note that the vertex set of $D(\mathcal{K}_k(\mathcal{A})\cap\mathcal{A}_{w'}^w(k))$
    is
    $V(D(\mathcal{K}_k(\mathcal{A})\cap\mathcal{A}_{w'}^w(k)))=\mathcal{K}_{k-1}(\mathcal{A})\cap \mathcal{A}_{w'-1}^w(k-1)$.
Thus,
    we can pick a vertex $\mathbf{v}$ with weight $w'-1$ in the form of
    $\mathbf{v}=(a_0,a_1,\ldots, a_{w'-2}, b_{w'-1},\ldots b_{k-2})$,
    where $a_0,a_1\ldots, a_{w'-2}\in\mathcal{W}$
    and $b_{w'-1},\ldots,b_{k-2}\in\mathcal{X}$.
Since $\mathbf{v}$ has weight $w'-1$,
    its predecessor must start with a weight-$1$ symbol.
Then note that $\mathbf{v}$ starts with $a_0\in\mathcal{W}$
    and
    no vertex can have two adjacent identical symbol,
    $\mathbf{v}$ has $|\mathcal{W}|-1$ predecessors.
Similarly,
    its successor must end with a weight-$1$ symbol,
    but since $\mathbf{v}$ ends with $b_{k-2}\in\mathcal{X}$,
    $\mathbf{v}$ has $|\mathcal{W}|$ successors.
Therefore,
    the in-degree of $\mathbf{v}$ is not equal to its out-degree,
    which implies that $D(\mathcal{K}_k(\mathcal{A})\cap\mathcal{A}_{w'}^w(k))$
    is not Eulerian.

The case $w\leq k-2$ is handled similarly.
Since Condition 1 fails to hold and $w'\leq w$,
    we can assume $w\geq 1$.
Then we can pick a vertex $\mathbf{v}'$ with weight $w$ in the form of
    $\mathbf{v}'=(a_0,a_1,\ldots, a_{w-1},b_{w},\ldots, b_{k-2})$,
    where $a_0,a_1\ldots, a_{w-1}\in\mathcal{W}$
    and $b_{w},\ldots,b_{k-2}\in\mathcal{X}$.
A similar argument shows that $\mathbf{v}'$ has
    in-degree $|\mathcal{X}|$ and out-degree $|\mathcal{X}|-1$,
    which are not equal.
\end{IEEEproof}
\begin{example}
Let
    $\mathcal{A}=\{A,T,C,G\}$
    and $k=3$.
First consider $w'=w=1$,
    which fails to meet
    any of the three conditions in Proposition \ref{prop:fixed_w_Kautz_exist}.
Fig.~\ref{fig:Kautz_fixed_wgt}\subref{subfig:Kautz_fixed_wgt_4_3_1_1}
    shows
    the fixed-weight Kautz graph
    $D(\mathcal{K}_3(\mathcal{A})\cap\mathcal{A}_{1}^1(3))$.
It can be seen that
    $D(\mathcal{K}_3(\mathcal{A})\cap\mathcal{A}_{1}^1(3))$ is not Eulerian
    since
    the vertex $CA$ has one successor $AT$
    and two predecessors, $AC$ and $TC$.
As a result,
    no fixed-weight Kautz sequence
    with parameters $(|\mathcal{A}|,k,w',w)=(4,3,1,1)$ exists.
This observation is consistent with Proposition \ref{prop:fixed_w_Kautz_exist}.

Next,
    consider the case $w'=1$ and $w=2$,
    which satisfies Condition 3 in 
    Proposition~\ref{prop:fixed_w_Kautz_exist}.
The fixed-weight Kautz graph
    $D(\mathcal{K}_3(\mathcal{A})\cap\mathcal{A}_{1}^2(3))$
    is shown in 
    Fig.~\ref{fig:Kautz_fixed_wgt}\subref{subfig:Kautz_fixed_wgt_4_3_1_2}.
It can be verified that
    the sequence
    $\mathbf{s}=CAGATCATGACACTACGAGTAGCTCTGTCGTG$
    can be represented by
    an Eulerian circuit on $D(\mathcal{K}_3(\mathcal{A})\cap\mathcal{A}_{1}^2(3))$.
Thus $\mathbf{s}$ is a fixed-weight Kautz sequence
    with parameters
    $(|\mathcal{A}|,k,w',w)=(4,3,1,2)$.
This result agrees with Proposition \ref{prop:fixed_w_Kautz_exist}.
\begin{figure}
    \centering
    \subfloat[][]{
        \scalebox{0.49}{
            \ifdefined\DrawTikz
                \begin{tikzpicture}[scale=1]

\tikzset{vertex/.style = {draw, circle, color=black, fill=black!5, line width=0.1cm,minimum size=2cm}}
\tikzset{arcA/.style = {->, line width=0.1cm, color = red!70!black}}
\tikzset{arcT/.style = {->, line width=0.1cm, color = green!70!black}}
\tikzset{arcC/.style = {->, line width=0.1cm, color = blue!70!black}}
\tikzset{arcG/.style = {->, line width=0.1cm, color = yellow!70!black}}

\node[vertex] (AT) at (-4,4) {\huge AT};
\node[vertex] (AC) at (5.656,0) {\huge AC};
\node[vertex] (AG) at (2,2) {\huge AG};
\node[vertex] (TA) at (-2,2) {\huge TA};
\node[vertex] (TC) at (-4,-4) {\huge TC};
\node[vertex] (TG) at (0,5.656) {\huge TG};
\node[vertex] (CA) at (-5.656,0) {\huge CA};
\node[vertex] (CT) at (-2,-2) {\huge CT};
\node[vertex] (GA) at (4,4) {\huge GA};
\node[vertex] (GT) at (0,-5.656) {\huge GT};

\draw[arcT] (CA) to [bend left=15] (AT);
\draw[arcT] (GA) to [bend left=15] (AT);
\draw[arcC] (TA) to [bend right=10] (AC);
\draw[arcG] (TA) to [bend left=15] (AG);

\draw[arcA] (CT) to [bend left=15] (TA);
\draw[arcA] (GT) to [bend right=10] (TA);
\draw[arcC] (AT) to [bend left=15] (TC);
\draw[arcG] (AT) to [bend left=15] (TG);

\draw[arcA] (AC) to [bend left=15] (CA);
\draw[arcA] (TC) to [bend left=15] (CA);
\draw[arcT] (AC) to [bend right=10] (CT);
\draw[arcT] (TC) to [bend left=15] (CT);

\draw[arcA] (AG) to [bend left=15] (GA);
\draw[arcA] (TG) to [bend left=15] (GA);
\draw[arcT] (AG) to [bend right=10] (GT);
\draw[arcT] (TG) to [bend left=15] (GT);


\end{tikzpicture}%
            \fi
        }
        \label{subfig:Kautz_fixed_wgt_4_3_1_1}
    }\\
    \subfloat[][]{
        \scalebox{0.49}{
            \ifdefined\DrawTikz
                \begin{tikzpicture}[scale=1]

\tikzset{vertex/.style = {draw, circle, color=black, fill=black!5, line width=0.1cm,minimum size=2cm}}
\tikzset{arcA/.style = {->, line width=0.1cm, color = red!70!black}}
\tikzset{arcT/.style = {->, line width=0.1cm, color = green!70!black}}
\tikzset{arcC/.style = {->, line width=0.1cm, color = blue!70!black}}
\tikzset{arcG/.style = {->, line width=0.1cm, color = yellow!70!black}}

\node[vertex] (AT) at (-4,4) {\huge AT};
\node[vertex] (AC) at (5.656,0) {\huge AC};
\node[vertex] (AG) at (2,2) {\huge AG};
\node[vertex] (TA) at (-2,2) {\huge TA};
\node[vertex] (TC) at (-4,-4) {\huge TC};
\node[vertex] (TG) at (0,5.656) {\huge TG};
\node[vertex] (CA) at (-5.656,0) {\huge CA};
\node[vertex] (CT) at (-2,-2) {\huge CT};
\node[vertex] (CG) at (4,-4) {\huge CG};
\node[vertex] (GA) at (4,4) {\huge GA};
\node[vertex] (GT) at (0,-5.656) {\huge GT};
\node[vertex] (GC) at (2,-2) {\huge GC};

\draw[arcT] (CA) to [bend left=15] (AT);
\draw[arcT] (GA) to [bend left=15] (AT);
\draw[arcC] (CA) to [bend left=15] (AC);
\draw[arcC] (TA) to [bend right=10] (AC);
\draw[arcC] (GA) to [bend left=15] (AC);
\draw[arcG] (TA) to [bend left=15] (AG);
\draw[arcG] (CA) to [bend right=10] (AG);
\draw[arcG] (GA) to [bend left=15] (AG);

\draw[arcA] (CT) to [bend left=15] (TA);
\draw[arcA] (GT) to [bend right=10] (TA);
\draw[arcC] (AT) to [bend left=15] (TC);
\draw[arcC] (CT) to [bend left=15] (TC);
\draw[arcC] (GT) to [bend left=15] (TC);
\draw[arcG] (AT) to [bend left=15] (TG);
\draw[arcG] (CT) to [bend right=10] (TG);
\draw[arcG] (GT) to [bend left=15] (TG);

\draw[arcA] (AC) to [bend left=15] (CA);
\draw[arcA] (TC) to [bend left=15] (CA);
\draw[arcA] (GC) to [bend right=10] (CA);
\draw[arcT] (AC) to [bend right=10] (CT);
\draw[arcT] (TC) to [bend left=15] (CT);
\draw[arcT] (GC) to [bend left=15] (CT);
\draw[arcG] (AC) to [bend left=15] (CG);
\draw[arcG] (TC) to [bend left=15] (CG);

\draw[arcA] (AG) to [bend left=15] (GA);
\draw[arcA] (CG) to [bend left=15] (GA);
\draw[arcA] (TG) to [bend left=15] (GA);
\draw[arcT] (AG) to [bend right=10] (GT);
\draw[arcT] (CG) to [bend left=15] (GT);
\draw[arcT] (TG) to [bend left=15] (GT);
\draw[arcC] (AG) to [bend left=15] (GC);
\draw[arcC] (TG) to [bend right=10] (GC);


\end{tikzpicture}%
            \fi
        }
        \label{subfig:Kautz_fixed_wgt_4_3_1_2}
    }
    \caption{
        Some fixed-weight Kautz graphs
            with alphabet $\mathcal{A}=\{A,T,C,G\}$ and $k=3$.
        \protect\subref{subfig:Kautz_fixed_wgt_4_3_1_1}
            The fixed-weight Kautz graph
            $D(\mathcal{K}_3(\mathcal{A})\cap\mathcal{A}_{1}^1(3))$.
        \protect\subref{subfig:Kautz_fixed_wgt_4_3_1_2}
            The fixed-weight Kautz graph
            $D(\mathcal{K}_3(\mathcal{A})\cap\mathcal{A}_{1}^2(3))$.
        }
    \label{fig:Kautz_fixed_wgt}
\end{figure}
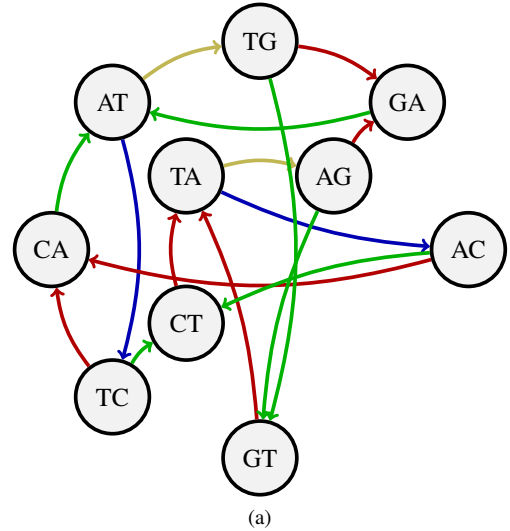
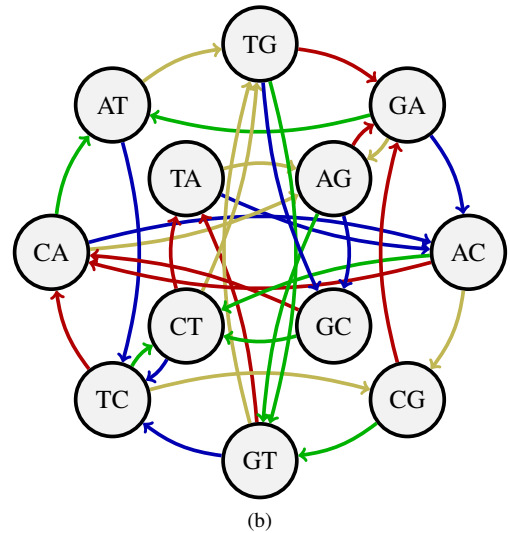
\end{example}

Orthogonality of fixed-weight Kautz sequences
    with $(w',w)=(0,0)$, $(k,k)$, or $(0,k)$
    can be deduced from the orthogonality of ordinary Kautz sequences, as explained in~\cite{Ortho_dB_Skiena}.
For other allowed cases of $(w',w)$ in Proposition \ref{prop:fixed_w_Kautz_exist},
    we have the following results.
\begin{proposition}\label{prop:fixed_w_Kautz_ortho}
Assume that $|\mathcal{W}|\geq 2$ and $|\mathcal{X}|\geq 2$.
Let $\delta\coloneqq |\mathcal{W}|+|\mathcal{X}|-1$
    denote the minimum in-out-degree of the ordinary Kautz graph $G^{\textnormal{Kautz}}_{\sigma,k}$.
We have the following:
\begin{itemize}
    \item 
        $D(\mathcal{K}_k(\mathcal{A})\cap\mathcal{A}_{1}^{k-1}(k))$ 
        has
        $\min(|\mathcal{W}|,|\mathcal{X}|,\lfloor \delta/2 \rfloor)$
        pairwise compatible Eulerian circuits.
    \item 
        $D(\mathcal{K}_k(\mathcal{A})\cap\mathcal{A}_{0}^{k-1}(k))$ 
        has
        $\min(|\mathcal{X}|,\lfloor \delta/2 \rfloor)$
        pairwise compatible Eulerian circuits.
    \item 
        $D(\mathcal{K}_k(\mathcal{A})\cap\mathcal{A}_{1}^{k}(k))$ 
        has
        $\min(|\mathcal{W}|,\lfloor \delta/2 \rfloor)$
        pairwise compatible Eulerian circuits.
\end{itemize}
\end{proposition}
\begin{IEEEproof}[Skecth of the proof]
We only prove the result for 
    the case of 
    $D(\mathcal{K}_k(\mathcal{A})\cap\mathcal{A}_{1}^{k-1}(k))$,
    since the other two cases can be handled similarly.

This proof is similar to that of Theorem \ref{thm:restr_ortho_deB}. Write $G\coloneqq D(\mathcal{K}_k(\mathcal{A})\cap\mathcal{A}_{1}^{k-1}(k))$
    for simplicity.
The vertex set of $G$
    is $V(G)=\mathcal{K}_{k-1}(\mathcal{A})$.
Any vertex
    in $G$
    with weight $0$ has in-out-degree $|\mathcal{W}|$,
    the ones with weight $k-1$ has in-out-degree $|\mathcal{X}|$, 
    and the other ones
    has in-out-degree $\delta$.
We apply the same vertex-splitting technique in the proof of Theorem \ref{thm:restr_ortho_deB}.
For each vertex $\mathbf{v}$ with weight $0$ or $k-1$,
    there exist
    $\min(|\mathcal{W}|,|\mathcal{X}|)$
    edge-disjoint wirings $W_1(\mathbf{v}),\ldots, W_{\min(|\mathcal{W}|,|\mathcal{X}|)}(\mathbf{v})$
    of $\mathbf{v}$.
Define the split graphs
    $G_1,\ldots, G_{\min(|\mathcal{W}|,|\mathcal{X}|)}$
    as follows:
$G_j$ is obtained by
    splitting each vertex $\mathbf{v}$ with weight $0$ or $k-1$
    along its $j$th wiring $W_j(\mathbf{v})$.
Then,
    as long as each split graph $G_j$ is Eulerian,
    we can select $\min(|\mathcal{W}|,|\mathcal{X}|,\lfloor\delta/2\rfloor)$ of them,
    merge each one back,
    and then rewire each remaining vertex with weight in $[1,k-2]$ one by one via
    Lemma \ref{lem:rewiring_general}.
This process gives $\min(|\mathcal{W}|,|\mathcal{X}|,\lfloor\delta/2\rfloor)$
    pairwise compatible Eulerian circuits.

To show that each split graph $G_j$ is Eulerian,
    first note that
    each split vertex has in-out-degree $1$
    and each unmodified vertex has in-out-degree $\delta$.
Thus each vertex in $G_j$ has equal in-degree and out-degree.
Then,
    we show that $G_j$ is strongly connected.
Note that
    it suffices to show that
    the subgraph $H$ of $G_j$ obtained by removing all the split vertices is strongly connected,
    since each split vertex
    can be either forwardly or reversely traversed
    to one vertex in $H$.
It is noteworthy that
    $H$ is the same for all $G_j$'s
    and 
    is identical to the subgraph of $G^{\textnormal{Kautz}}_{\sigma,k}$
    obtained by removing all the vertices with weight $0$ or $k-1$.

Let $\mathbf{v}=(v_0,\ldots,v_{k-2})$ and $\mathbf{v}'=(v'_0,\ldots,v'_{k-2})$
    be two vertices in $H$.
Since the weight of $\mathbf{v}$ is at least $1$,
    there must be a symbol in $\mathbf{v}$ that is weighted.
That is,
    there exists some $t\in[0,k-2]$ such that $v_t\in \mathcal{W}$.
Now consider the
    walk
    $P$
    on $G^{\textnormal{Kautz}}_{\sigma,k}$ from $\mathbf{v}$ to $\hat{\mathbf{v}}=(v_{t+1},\ldots,v_{k-2},\hat{v}_0,\ldots,\hat{v}_{t})$
    that induces the sequence
    $(v_0,\ldots, v_{k-2},\hat{v}_0,\ldots, \hat{v}_{t})$,
    where $\hat{v}_0,\ldots, \hat{v}_{t}$ are
    defined as follows:
\begin{itemize}
    \item
        If $v_{k-2}\neq v_0$, then choose $\hat{v}_s=v_s$ for $s\in[0,t]$.
    \item 
        If $v_{k-2}=v_0$, first choose a permutation $\pi$ on
            the set $v_0$ belongs to ($\mathcal{W}$ or $\mathcal{X}$)
            such that $v_0$ is not fixed by $\pi$.
        Then for $s\in[0,t]$, define
            \begin{align*}
                \hat{v}_s \coloneqq
                \begin{cases}
                    \pi(v_s), \textnormal{ if }v_s\textnormal{ and }v_0 \textnormal{ belong to the same set},\\
                    v_s, \textnormal{ if }v_s\textnormal{ and }v_0 \textnormal{ belong to different  sets}.\\
                \end{cases}
            \end{align*}
\end{itemize}
It can be seen that
    the weight representation of the sequence
    $(v_0,\ldots, v_{k-2},\hat{v}_0,\ldots, \hat{v}_{t})$
    is the same as 
    $(v_0,\cdots, v_{k-2},v_0,\ldots, v_{t})$.
This implies that
    the walk
    $P$
    only visits vertices having the same weight as $\mathbf{v}$.
Therefore,
    $P$ is also a valid walk in the subgraph $H$.

Similarly,
    since the weight of $\mathbf{v}'$ is at most $k-2$,
    there exists some $u\in[0,k-2]$ such that $v'_u\in \mathcal{X}$.
Then,
    consider the vertex
    $\hat{\mathbf{v}}'=(\hat{v}'_u,\ldots,\hat{v}'_{k-2},v'_0,\ldots,v'_{u-1})$,
    where
    $\hat{v}'_u,\ldots,\hat{v}'_{k-2}$
    are defined in a similar way as follows:
\begin{itemize}
    \item
        If $v'_0\neq v'_{k-2}$, define $\hat{v}'_r=v'_r$
            for $r\in[u,k-2]$.
    \item 
        If $v'_0=v'_{k-2}$,
            find a permutation $\pi'$ on the set
            $v'_0$ belongs to
            that does not fix $v'_0$.
        Then
            for $r\in[u,k-2]$
            we define $\hat{v}'_r\coloneqq \pi'(v'_r)$
            if $v'_r$ and $v'_0$ belongs to the same set
            and $\hat{v}'_r\coloneqq v'_r$ otherwise.
\end{itemize}
It follows that
    the walk
    from $\hat{\mathbf{v}}'$ to $\mathbf{v}'$
    that 
    induces the sequence
    $(\hat{v}'_u,\ldots, \hat{v}'_{k-2}, v'_0,\ldots v'_{k-2})$
    is a valid walk in $H$.
Then,
    note that $\hat{v}_t$ belongs to the same set as $v_t$,
    which is $\mathcal{W}$.
Similarly,
    $\hat{v}_u'\in\mathcal{X}$.
In particular,
    we have $\hat{v}_t\neq \hat{v}_u'$.
Thus,
    we can walk in $G^{\textnormal{Kautz}}_{\sigma,k}$
    from $\hat{\mathbf{v}}$ to $\hat{\mathbf{v}}'$
    inducing the sequence
    $(v_{t+1},\ldots,v_{k-2},\hat{v}_1,\ldots, \hat{v}_t,
    \hat{v}_u',\ldots,\hat{v}_{k-2}', v'_0 \ldots, v'_{u-1})$.
Denote this walk as $\bar{P}$.
Furthermore,
    any vertex visited by
    $\bar{P}$
    contains the substring $\hat{v}_t\hat{v}_u'$.
Recall that $\hat{v}_t$ has weight $1$
    and $\hat{v}_u'$ has weight $0$.
As a result,
    any vertex visited by
    $\bar{P}$
    must have weight between $2$ and $k-2$,
    implying that
    the whole walk $\bar{P}$
    lies in $H$.
In conclusion,
    we can walk in $H$ from $\mathbf{v}$ to $\hat{\mathbf{v}}$,
    then from $\hat{\mathbf{v}}$ to $\hat{\mathbf{v}}'$,
    and then from $\hat{\mathbf{v}}'$ to $\mathbf{v}'$.
These arguments demonstrate  that $H$,
    and thus each split graph $G_j$,
    is strongly connected.

These arguments prove the statement of the result.
\end{IEEEproof}


\textbf{Acknowledgment:} The work was supported in part by the NSF grant number 2107344.

\bibliographystyle{IEEEbib.bst}
\bibliography{refs}

\end{document}